\documentclass[twoside]{article}

\usepackage[preprint]{aistats2027}

\usepackage{amsmath,amssymb,amsfonts,amsthm,mathtools,bm}
\usepackage{booktabs,array}
\usepackage{graphicx}
\usepackage{microtype}
\usepackage{xcolor}
\usepackage{url}
\usepackage{hyperref}
\usepackage{subcaption}
\usepackage{enumitem}
\usepackage{cleveref}
\usepackage{wrapfig}

\usepackage[round]{natbib}
\renewcommand{\bibname}{References}
\renewcommand{\bibsection}{\subsubsection*{\bibname}}

\newtheorem{theorem}{Theorem}
\newtheorem{lemma}{Lemma}
\newtheorem{corollary}{Corollary}
\newtheorem{proposition}{Proposition}

\DeclareMathOperator*{\argmax}{arg\,max}

\newcommand{\E}{\mathbb{E}}
\newcommand{\Prob}{\mathbb{P}}
\newcommand{\1}{\mathbf{1}}
\newcommand{\cF}{\mathcal{F}}
\newcommand{\eps}{\epsilon}
\newcommand{\mS}{m^\star}
\newcommand{\vpi}{\boldsymbol v}
\newcommand{\rb}{\boldsymbol r}

\newcommand{\Info}{\mathcal{I}}
\newcommand{\J}{\mathsf{J}}
\newcommand{\prior}{\pi_0}
\newcommand{\Kprior}{\mathcal{K}_{\prior}}

\begin{document}

\twocolumn[
\aistatstitle{Pathwise Information Certificates for Decentralized Adaptive Sensing}
\aistatsauthor{ Theodoros Tsiligkaridis }
\aistatsaddress{ MIT Lincoln Laboratory }
]

\begin{abstract}
We study decentralized adaptive sensing, where multiple agents choose measurements from evolving local beliefs while exchanging information over a communication graph. We ask whether the measurements actually selected by an adaptive policy have collected enough evidence to distinguish the true target from every plausible alternative. We develop a pathwise certificate based on the R\'enyi--Chernoff information accumulated along the realized sensing trajectory. It yields nonasymptotic MAP-error bounds and an anytime, network-wide stopping rule for arbitrary history-dependent sensing policies, while separating accumulated statistical information from a bounded network-mixing transient. Linear growth of the information against the least-resolved competitor implies exponential decay of MAP and squared-localization error. A classical pairwise KL converse, specialized to the adaptive decentralized transcript, shows that insufficient information on any pair prevents a positive uniform error exponent, confirming the hardest competitor as a fundamental bottleneck. Across policies, graph topologies, sensor profiles, and seeds, the worst-competitor score correlates more strongly with localization speed than an average-pair proxy in both 1D ($r=0.89$ versus $0.40$) and structured 2D sensing ($r=0.77$ versus $0.48$). Our results provide a practical way to certify and diagnose adaptive multi-agent sensing systems using the evidence they actually collect.
\end{abstract}

\section{Introduction}
\label{sec:introduction}

Adaptive information acquisition arises when sensing is itself a decision, from active perception and environmental monitoring to sequential diagnosis \citep{bajcsy2018active,singh2009informative,gorry1968sequential}. In a decentralized system, performance depends on both \emph{which evidence is acquired} and \emph{how that evidence propagates through the network}. The key difficulty arises when sensing decisions depend on evolving local beliefs: the observation kernel changes with the realized history, so a useful reliability guarantee should remain valid for data-dependent policies and stopping times.

Existing theories address important pieces of this problem separately. Distributed social-learning analyses typically assume fixed local observation laws and quantify how statistical information and graph topology drive learning \citep{nedic2017fast,lalitha2018social,shahrampour2016distributed,wu2025frequentist}. Controlled sensing and active testing analyze designed action policies, action frequencies, error exponents, or sequential tests \citep{chernoff1959sequential,nitinawarat2013controlled,lalitha2017active,rangi2021distributed,hsu2025tradeoffs}. Classical testing already identifies hard alternatives as rate
bottlenecks; our focus is different: given the trajectory produced by
an adaptive decentralized policy, we ask whether the actions it
\emph{actually selected} accumulated enough evidence against every
alternative to certify reliable localization.

\begin{figure}[t]
\centering
\includegraphics[width=0.98\columnwidth]{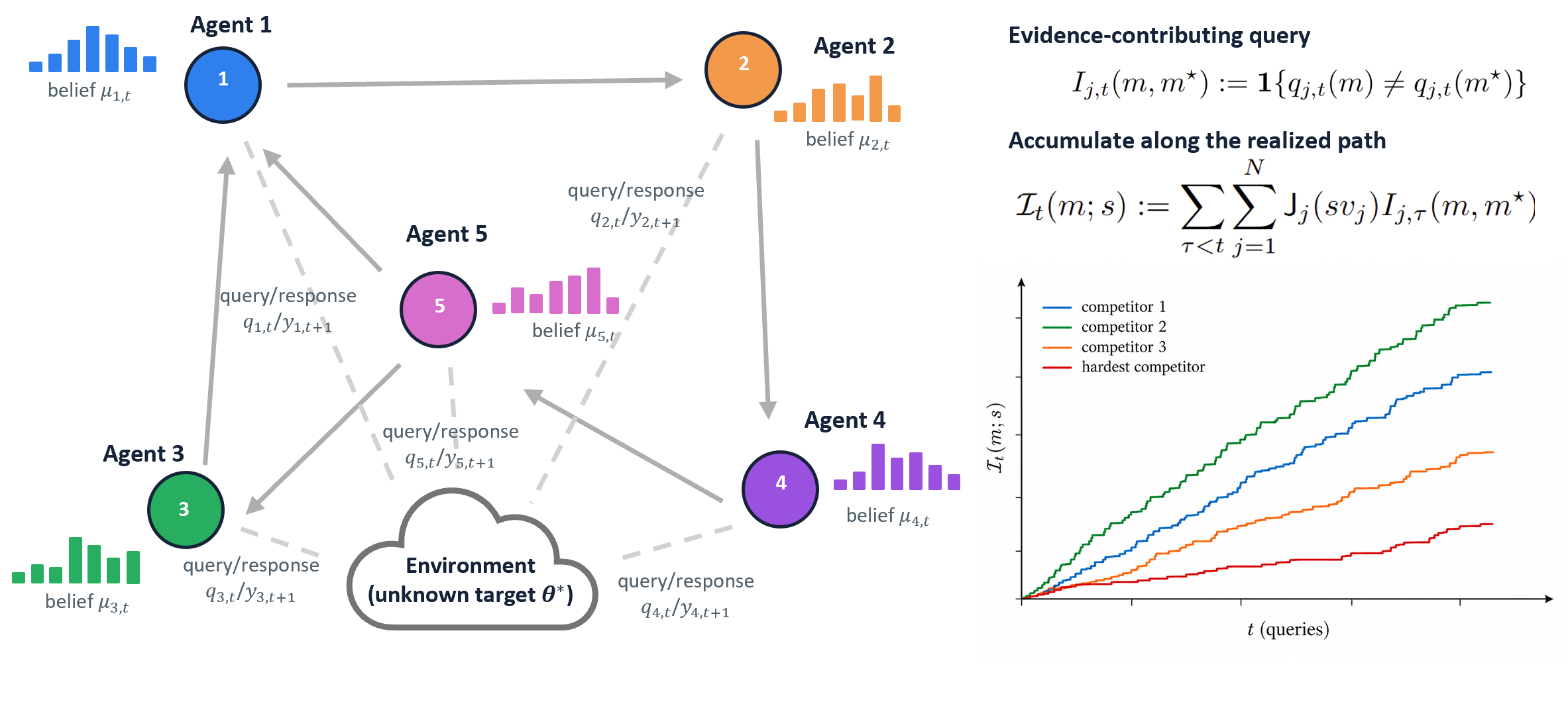}
\caption{\textbf{Decentralized adaptive sensing and the pathwise certificate.}
Agents select history-dependent queries from evolving local beliefs, observe
noisy responses, and pool beliefs over a communication graph. Each separating
query contributes pairwise Chernoff information $\J_j(sv_j)$, which depends on the sensor model and the agent's stationary network influence. Accumulating this evidence along the \emph{realized} trajectory yields the information process $\Info_t(m;s)$, which drives finite-time and anytime reliability guarantees against the hardest
competitor.}
\label{fig:overview}
\end{figure}

Our certificate accumulates pairwise Chernoff information from the
queries actually selected. A measurement contributes only to hypothesis
pairs that it separates, with the size of that contribution determined
jointly by the sensor model and stationary network. The least accumulated evidence over competing hypotheses controls the finite-time and anytime guarantees, while network disagreement appears as a bounded transient (Figure~\ref{fig:overview}).

The construction is not tied to a particular sensing policy or target geometry. It evaluates the information induced by the \emph{realized actions} of an arbitrary history-dependent policy, so it can be applied post hoc to heuristic, posterior-driven, or learned sensing policies without changing how those policies are trained. We use binary queries because they make pairwise distinguishability especially transparent, and illustrate the theory with ordered 1D thresholds and a structurally
different 2D half-plane library. Appendix~\ref{app:general-observation} extends the same construction to bounded action-dependent observation kernels beyond the binary symmetric channel model.

\paragraph{Contributions.}
\begin{enumerate}[leftmargin=*,itemsep=1pt,topsep=2pt]
\item \textbf{Pathwise finite-time and anytime reliability.} We accumulate pairwise Chernoff information along the realized adaptive trajectory, with each separating measurement contributing an amount that depends on its sensor model and stationary graph influence. This yields node-wise finite-horizon MAP bounds and a network-wide anytime certificate for arbitrary history-dependent policies, with network disagreement appearing only as a bounded transient.

\item \textbf{Hardest-competitor rates with a classical converse.} Linear growth of the least-resolved competitor's realized information yields exponential MAP and squared-localization-loss rates. The hard-pair principle and two-point lower-bound machinery are classical; our contribution is to connect them to the same realized adaptive decentralized transcript used by the certificate, and to show why average pair coverage can hide an unresolved alternative.

\item \textbf{Policy-agnostic certification with optional guaranteed coverage.} The results apply to randomized, heuristic, posterior-driven, or learned policies. A certified-exploration wrapper separately guarantees worst-pair coverage while leaving all other rounds to the base policy.

\item \textbf{Empirical tests in 1D and 2D.} Across policies, graphs, sensors, and seeds, we test exponential localization, graph effects, and the hardest-competitor hypothesis in ordered 1D and structured 2D. In both, worst-competitor information is more strongly associated with localization speed than an average-pair proxy.
\end{enumerate}

\section{Related Work}\label{sec:related}

\paragraph{Adaptive querying and controlled sensing.}
Noisy twenty questions and probabilistic bisection adapt queries to reduce uncertainty \citep{horstein1963sequential,jedynak2012twenty,waeber2013bisection,chung2018uep,zhou2021adaptive}; decentralized variants combine active queries with belief sharing \citep{tsiligkaridis2014collaborative,tsiligkaridis2015decentralized,tsiligkaridis2017correction,tsiligkaridis2017distributed,tsiligkaridis2018adaptive}. Controlled sensing instead studies policy design and error exponents \citep{chernoff1959sequential,nitinawarat2013controlled,hsu2025tradeoffs}. We certify the evidence collected by a \emph{realized} adaptive trajectory rather than optimize a policy class.

\paragraph{Decentralized learning and active testing over networks.}
Log-linear social learning gives consistency and finite-time or topology-dependent guarantees under fixed observation models \citep{jadbabaie2012nonbayesian,rahimian2015loglinear,nedic2017fast,lalitha2018social,shahrampour2016distributed,wu2025frequentist}. Active network testing permits selected sensing actions \citep{lalitha2017active,rangi2021distributed}, including learned policies \citep{szostak2024deep}. Prior combinations of decentralization and active sensing analyze specific strategies or asymptotic characteristics; ours conditions on the complete realized action trajectory and is nonasymptotic and anytime-valid (Table~\ref{tab:positioning}).

\begin{table*}[t]
\centering
\caption{\textbf{Positioning of our work.} 
The novelty of our guarantee is the conjunction of decentralization and adaptive sensing with a finite-time, anytime-valid certificate for the \emph{realized trajectory} of an otherwise arbitrary policy.}
\label{tab:positioning}
\scriptsize
\setlength{\tabcolsep}{3.5pt}
\begin{tabular}{lcccccc}
\toprule
Work & Decentralized & Adaptive sensing & Arbitrary-policy cert. & Realized-path cert. & Finite-time & Anytime stop \\
\midrule
\citet{nedic2017fast}                  & $\checkmark$ & --            & -- & -- & $\checkmark$ & -- \\
\citet{nitinawarat2013controlled}      & --            & $\checkmark$ & -- & -- & -- & -- \\
\citet{tsiligkaridis2015decentralized} & $\checkmark$ & $\checkmark$ & -- & -- & -- & -- \\
\citet{lalitha2017active}              & $\checkmark$ & $\checkmark$ & -- & -- & -- & -- \\
\citet{rangi2021distributed}           & $\checkmark$ & $\checkmark$ & -- & -- & -- & -- \\
\textbf{This work}                     & $\checkmark$ & $\checkmark$ & $\checkmark$ & $\checkmark$ & $\checkmark$ & $\checkmark$ \\
\bottomrule
\end{tabular}
\end{table*}

To our knowledge, prior work does not jointly provide finite-time and
anytime-valid guarantees that certify whether the measurements actually
selected by an arbitrary decentralized, history-dependent sensing policy
have accumulated sufficient evidence for reliable localization.


\section{Decentralized Adaptive Querying}\label{sec:model}
Let $\Theta=\{\theta_1,\ldots,\theta_M\}$ be a finite hypothesis set and let $\mS\in[M]$ denote the unknown target. For localization, $\theta_m$ may lie in any metric space, including $\mathbb R^d$; the theory uses only the induced hypothesis-indexed observation laws. Agent $j$ selects $q_{j,t}:[M]\to\{0,1\}$ from its local history $\cF^{\rm loc}_{j,t}$. For analysis, $\cF_t$ denotes the global filtration after the round-$t$ queries (and their policy randomization) are selected but before the next observations, with $\cF^{\rm loc}_{j,t}\subseteq\cF_t$. Then
\begin{align}
Y_{j,t+1}&=q_{j,t}(\mS)\oplus W_{j,t+1},\nonumber\\
W_{j,t+1}&\sim\mathrm{Bernoulli}(\eps_j),\qquad 0<\eps_j<\tfrac12,
\end{align}
and conditional on $\cF_t$ the channel noises are independent across agents and time. The local likelihood used in the Bayes update is
\begin{equation}
\ell_{j,t}(y\mid m)=
\begin{cases}
1-\eps_j, & y=q_{j,t}(m),\\
\eps_j, & y\neq q_{j,t}(m),
\end{cases}
\qquad y\in\{0,1\}.
\label{eq:local-likelihood}
\end{equation}

Let $\mu_{i,t}(m)$ denote agent $i$'s posterior belief in hypothesis $m$ at round $t$. Agents start from a common full-support prior $\mu_{i,0}=\prior$. Let $A\in\mathbb R^{N\times N}$ be primitive (irreducible and aperiodic) and row stochastic, with stationary left eigenvector $\vpi^\top A=\vpi^\top$, $\vpi>0$, $\vpi^\top\mathbf1=1$. Each round consists of a local Bayes update followed by log-linear pooling:
\begin{align}
\widetilde\mu_{j,t+1}(m)
&=\frac{\mu_{j,t}(m)\,\ell_{j,t}(Y_{j,t+1}\mid m)}
{\sum_u\mu_{j,t}(u)\,\ell_{j,t}(Y_{j,t+1}\mid u)},\nonumber\\
\mu_{i,t+1}(m)
&=\frac{\prod_j\widetilde\mu_{j,t+1}(m)^{A_{ij}}}
{\sum_u\prod_j\widetilde\mu_{j,t+1}(u)^{A_{ij}}}.
\label{eq:pool}
\end{align}
Thus each round has a simple “local evidence, then consensus” structure: each agent first performs a Bayesian update and then averages log-beliefs through the communication network. This geometric pooling rule is standard in distributed non-Bayesian/social learning \citep{jadbabaie2012nonbayesian,rahimian2015loglinear,nedic2017fast,lalitha2018social}. Agent $i$ predicts 
\begin{equation*}
    \widehat m_{i,t}\in\argmax_m\mu_{i,t}(m).
\end{equation*}

\paragraph{Query policies.} In the experiments, we evaluate posterior bisection, Thompson-style sampling, random querying, and exploration+bisection in 1D and 2D. Exact policy definitions are given in Appendix~\ref{app:experimental-details}. The theory does not assume any of these policies.

\paragraph{Complexity.}
The Bayes update costs $O(M)$ per agent and pooling costs $O(M|E|)$ network-wide per round. Ordered-threshold selection is $O(M)$; naive scoring of a finite library $\mathcal Q$ is $O(M|\mathcal Q|)$ per agent. Explicit anytime tracking additionally stores $O(M^2)$ pair scores and costs $O(NM^2)$ per round naively; structured query families can reduce this overhead. We implement belief updates in log space.

\section{Reliability Certificates from Realized Information}\label{sec:theory}

\paragraph{Notation and accumulated evidence.}
Fix a true target $\mS$ and a false target $m\neq\mS$. Two things determine whether a measurement is useful for a pair of hypotheses: whether the query separates that pair, and how informative the corresponding sensor observation is. A query contributes evidence for this pair only when the two targets predict different noiseless answers, so define
\[
I_{j,t}(m,\mS):=\1\{q_{j,t}(m)\neq q_{j,t}(\mS)\}.
\]
For a separating query, agent $j$ must distinguish the two binary symmetric channel (BSC) response laws with crossover probability $\eps_j$. For $\alpha\in[0,1]$, define its pairwise information contribution
\begin{equation}
\J_j(\alpha):=-\log\!\left[(1-\eps_j)^{1-\alpha}\eps_j^\alpha+\eps_j^{1-\alpha}(1-\eps_j)^\alpha\right].
\label{eq:J-main}
\end{equation}
It is the negative log Chernoff coefficient (equivalently, a scaled R\'enyi divergence) between the two BSC output laws \citep{chernoff1952measure,renyi1961measures,vanerven2014renyi}; larger values mean that one separating observation is more informative. Let
\begin{align}
\mathcal S&:=\{s>0:sv_j\le1\ \forall j\},\nonumber\\
\Info_t(m;s)&:=\sum_{\tau<t}\sum_{j=1}^{N}\J_j(sv_j)I_{j,\tau}(m,\mS).
\label{eq:information}
\end{align}
The set $\mathcal S=(0,\,1/\max_j v_j]$ is always nonempty because $s=1$ is admissible. Relative to a balanced graph ($v_j=1/N$), strong stationary imbalance narrows this interval and can prevent the terms $sv_j$ from being simultaneously near their most informative values. Appendix~\ref{app:imbalance-proof} quantifies the resulting certified-information loss. $\Info_t(m;s)$ is the evidence accumulated against competitor $m$ along the realized queries, and $s$ is an analysis parameter rather than part of the sensing policy.
For the anytime guarantee $s$ is fixed in advance; only the post-hoc diagnostic optimizes it after a rollout. For the finite-time network correction, define
\begin{equation}
b_j:=\log\frac{1-\eps_j}{\eps_j},\qquad
\Delta_i:=\sum_{k\ge1}\sum_{j=1}^{N}|(A^k)_{ij}-v_j|b_j,
\label{eq:Delta}
\end{equation}
and the prior factor $\Kprior(\mS,s):=\sum_{m\neq\mS}(\prior(m)/\prior(\mS))^s$. 
For the network-wide anytime result below, we will use the corresponding worst-case quantities
\[
\Delta_{\max}:=\max_i\Delta_i,
\qquad
\mathcal K_{\max}(s):=
\max_{m'}\Kprior(m',s).
\]
For a uniform prior,
$\Kprior(\mS,s)=\mathcal K_{\max}(s)=M-1$.
Primitivity of $A$ implies $\Delta_i<\infty$ because $A^k$ converges
geometrically to $\mathbf 1\vpi^\top$
\citep{seneta2006nonnegative,levin2017markov}.

\paragraph{Offline-computable certificate constants.}
The admissible set $\mathcal S$, stationary weights $\vpi$, information
functions $\J_j$, mixing corrections $\Delta_i$ and $\Delta_{\max}$, and
prior factors $\Kprior$ and $\mathcal K_{\max}$ depend only on the known
prior, communication graph, and sensor models and can therefore be
precomputed before deployment. During operation, the anytime certificate
only needs to update the pairwise information accumulated by the realized
queries.

\subsection{Finite-time reliability}
For any chosen evidence threshold \(a_{m,t}\), the error probability is bounded by the probability that the policy has not yet accumulated that evidence plus an exponentially small residual due to noise and network disagreement.
\begin{theorem}[Finite-time adaptive-query localization]
\label{thm:main}
Fix agent $i$, true target $\mS$, $s\in\mathcal S$, and deterministic levels $a_{m,t}\ge0$. For any history-dependent query policy,
\begin{align}
\Prob_{\mS}(\widehat m_{i,t}\neq\mS)
\le\sum_{m\neq\mS}\Bigg[&\Prob_{\mS}\!\left(\Info_t(m;s)<a_{m,t}\right)\nonumber\\
&+\left(\frac{\prior(m)}{\prior(\mS)}\right)^s e^{s\Delta_i-a_{m,t}}\Bigg].
\label{eq:main-bound}
\end{align}
\end{theorem}
The first term measures whether the policy has collected enough evidence against $m$; the second is the residual effect of observation noise and network disagreement. Thus the bound separates \emph{information acquisition} from decentralized communication transients.

\paragraph{Proof idea.}
The full proof is in Appendix~\ref{app:proofs}. Projecting posterior log-odds onto $\vpi$ removes the repeated geometric-pooling dynamics. Conditional on the history, the selected queries are fixed, and the fractional moment of each likelihood ratio is exactly $\exp[-\J_j(sv_j)]$ when the query separates the pair and $1$ otherwise. Multiplying the projected odds by $\exp[\Info_t(m;s)]$ therefore cancels this conditional drift and produces a nonnegative martingale. The bounded difference between projected and node-wise log-odds contributes the factor $e^{s\Delta_i}$.

For geometrically mixing $A$, $\Delta_i$ is finite; Appendix~\ref{app:mixing-proof} gives explicit geometric and spectral bounds. Thus poorer mixing changes the finite-time offset rather than which competitors are informative.


\subsection{Anytime network-wide reliability} \label{sec:anytime}
Theorem \ref{thm:main} is truth-indexed and therefore most natural for fixed-horizon analysis. A truth-independent stopping certificate cannot depend on the unknown target, so we replace truth-versus-competitor quantities by the minimum evidence over all candidate pairs. For distinct targets $m,m'$, write
$I_{j,t}(m,m'):=\1\{q_{j,t}(m)\neq q_{j,t}(m')\}$ and define the counterparts
\begin{equation}
\begin{aligned}
\Info_t(m,m';s)&:=\sum_{\tau<t}\sum_j\J_j(sv_j)I_{j,\tau}(m,m'),\\
\underline{\Info}_t(s)&:=\min_{m\neq m'}\Info_t(m,m';s).
\end{aligned}
\label{eq:pairwise-information}
\end{equation}

\begin{theorem}[Anytime network-wide certificate]
\label{thm:anytime}
For fixed $s\in\mathcal S$ and $\delta\in(0,1)$, define
\begin{align}
c_\delta(s)&:=s\Delta_{\max}+\log\frac{N\mathcal K_{\max}(s)}{\delta},\nonumber\\
\tau_\delta(s)&:=\inf\{t:\underline{\Info}_t(s)\ge c_\delta(s)\}.
\label{eq:anytime-threshold}
\end{align}
Then, for every true target $\mS$,
\[
\Prob_{\mS}\!\left(\exists i,\ \exists t\ge\tau_\delta(s):\widehat m_{i,t}\neq\mS\right)\le\delta.
\]
\end{theorem}
The rule stops only after \emph{every} target pair has enough evidence, so validity does not depend on knowing the target or fixing a horizon. Computing $\underline{\Info}_t(s)$ requires the network-wide selected-query log: Eq.~\eqref{eq:pool} alone does not make it locally available at each node. Deployment therefore assumes either a centralized logger/monitor or an auxiliary dissemination/aggregation protocol for query identities; the theorem is agnostic to that protocol. If $\underline{\Info}_t(s)\ge\gamma t-B$, then
\[
\tau_\delta(s)\le
\left\lceil\frac{B+s\Delta_{\max}+\log(N\mathcal K_{\max}(s)/\delta)}{\gamma}\right\rceil.
\]
The fixed $s$ is a one-dimensional offline design choice: for certified exploration one may maximize $W(s)$ in Eq.~\eqref{eq:sweep-exp}; more generally, if a lower bound $\underline{\Info}_t(s)\ge\gamma(s)t-B(s)$ is available, one may minimize the displayed stopping-time bound over $s\in\mathcal S$ before monitoring begins. Data-dependent online tuning of $s$ would require a separate time-uniform argument.

\paragraph{Three uses.}
Theorem~\ref{thm:main} gives fixed-horizon bounds, Theorem~\ref{thm:anytime} gives a truth-independent stopping condition, and Eq.~\eqref{eq:empirical-certificate} gives a truth-aware post-hoc policy diagnostic. Only the last optimizes $s$ after the rollout.

\subsection{Exponential rates and why the hardest competitor matters}
Define the rate $E_i:=\liminf_{t\to\infty}-t^{-1}\log\Prob_{\mS}(\widehat m_{i,t}\neq\mS)$.
\begin{corollary}[Stochastic information growth implies exponential learning]
\label{cor:stochastic-growth}
If for some fixed $s\in\mathcal S$, $\gamma,\kappa>0$, finite $B$, and constants $C_m$,
\[
\Prob_{\mS}\!\left(\Info_t(m;s)<\gamma t-B\right)\le C_m e^{-\kappa t}
\]
for every false $m$ and all sufficiently large $t$, then $E_i\ge\min\{\gamma,\kappa\}$. If the information lower bound holds deterministically, $E_i\ge\gamma$.
\end{corollary}
Here $\gamma$ is the information-growth rate and $\kappa$ controls its lower-tail reliability; the slower mechanism limits the guaranteed exponent.

\begin{corollary}[MAP and squared localization loss]
\label{cor:mse}
Equip the finite target set with a metric $d$ and assume distinct targets, so $d_{\min}:=\min_{m\neq m'}d(\theta_m,\theta_{m'})>0$. With $D_\Theta:=\max_{m,m'}d(\theta_m,\theta_{m'})$,
\begin{align}
 d_{\min}^2\Prob_{\mS}(\widehat m_{i,t}\neq\mS)
 &\le \E_{\mS}[d(\widehat\theta_{i,t},\theta_{\mS})^2] \nonumber \\
 &\le D_\Theta^2\Prob_{\mS}(\widehat m_{i,t}\neq\mS).
\label{eq:mse-sandwich}
\end{align}
Hence MAP error and squared metric localization loss have the same exponent.
\end{corollary}

The hard-pair/minimax principle is classical \citep{chernoff1952measure,chernoff1959sequential,nitinawarat2013controlled,bretagnolle1979estimation,tsybakov2009introduction}. Here the unresolved pair is identified by the \emph{realized} decentralized action trajectory. The next results show that an average can hide such a pair and that pairwise distinguishability is necessary for a uniform exponent.

\begin{proposition}[Average pair coverage can hide localization failure]
\label{prop:aggregate-insufficient}
For any $M\ge4$ ordered targets with thresholds $q_k(m)=\1\{m\le k\}$, consider synchronized threshold schedules in which all agents use the same threshold on a given round. Let $k_\star=\lfloor M/2\rfloor$. Repeating only $q_{k_\star}$ separates
\begin{equation}
\phi_{\rm one}(M):=\frac{k_\star(M-k_\star)}{\binom{M}{2}}
\label{eq:single-cut-fraction}
\end{equation}
of all unordered pairs on every round, yet at least two hypotheses on the same side are never distinguished, so for every horizon $T$,
\[
\sup_{m^\star}\Prob_{m^\star}(\widehat m_T\neq m^\star)\ge\tfrac12.
\]
By contrast, a balanced cycle through all $M-1$ thresholds has worst-pair separation frequency $1/(M-1)$ and average pair-separation fraction
\begin{equation}
\phi_{\rm cyc}(M)=\frac{1}{(M-1)\binom{M}{2}}\sum_{k=1}^{M-1}k(M-k)=\frac{M+1}{3(M-1)}.
\label{eq:cycle-fraction}
\end{equation}
For every $M\ge4$, $\phi_{\rm one}(M)>\phi_{\rm cyc}(M)$, although the cyclic schedule repeatedly separates every pair and therefore admits a positive certified exponent under Proposition~\ref{prop:sweep}.
\end{proposition}
This is why our empirical diagnostic uses the minimum over competitors:
an average information score can look favorable even when one unresolved
alternative still dominates localization error. The average-pair quantity is used in the experiments as an ablation of the minimum-over-competitors. For $M=32$, Proposition~\ref{prop:aggregate-insufficient} gives the $0.516$ versus $0.355$ comparison used in the 1D experiment. Appendix~\ref{app:aggregate-proof} gives the proof.

\paragraph{Classical pairwise converse.}
Let $\mathsf P_m^T$ be the law of the complete adaptive transcript through round $T$ under hypothesis $m$. The following is a specialization of the standard Bretagnolle--Huber/Le Cam two-point argument \citep{bretagnolle1979estimation,tsybakov2009introduction}.
\begin{proposition}[Pairwise KL converse for the adaptive transcript]
\label{prop:pairwise-converse}
For any estimator,
\begin{equation}
\sup_{u\in[M]}\Prob_u(\widehat m_T\neq u)
\ge \frac14\exp\!\left[-\min_{m\neq m'}D_{\rm KL}(\mathsf P_m^T\|\mathsf P_{m'}^T)\right].
\label{eq:pairwise-converse}
\end{equation}
For the BSC query model and a common history-dependent policy kernel,
\begin{align}
D_{\rm KL}(\mathsf P_m^T\|\mathsf P_{m'}^T)
&=\E_m\!\left[\sum_{t<T}\sum_j d_j\,\1\{q_{j,t}(m)\neq q_{j,t}(m')\}\right],\nonumber\\
d_j&=(1-2\eps_j)\log\frac{1-\eps_j}{\eps_j}.
\label{eq:bsc-kl-converse}
\end{align}
Hence sublinear expected information for even one pair rules out a positive \emph{uniform} localization-error exponent.
\end{proposition}
Together with the finite-time upper bound, this gives a two-sided message:
resolving the hardest remaining competitor is sufficient for certification
and also necessary for any method to achieve a positive uniform error exponent.


\paragraph{Beyond binary observations.}
The proof extends beyond BSC likelihoods. For a general action-dependent observation kernel, Eq.~\eqref{eq:J-main} is replaced by the corresponding Chernoff coefficient between the two action-conditioned output laws. Under uniformly bounded pairwise log-likelihood ratios, the finite-time and anytime results remain valid; Appendix~\ref{app:general-observation} gives the construction and bounded continuous example.

\section{Certified Exploration for Arbitrary Adaptive Policies}\label{sec:rates}
The certificate diagnoses realized evidence but does not force pairwise exploration. To guarantee coverage while otherwise preserving the base policy, let $\mathcal Q_{\rm cert}=\{q^{(1)},\ldots,q^{(L)}\}$ be a synchronized finite separating family with minimum response-code distance
\begin{equation}
d_{\mathcal Q}:=\min_{m\neq m'}\sum_{\ell=1}^{L}
\1\{q^{(\ell)}(m)\neq q^{(\ell)}(m')\}.
\label{eq:cert-distance}
\end{equation}
Because $\mathcal Q_{\rm cert}$ is separating, $d_{\mathcal Q}\ge1$. With $W(s):=\sum_j\J_j(sv_j)$, after $C_t$ complete certified cycles every pair has received at least $C_td_{\mathcal Q}W(s)$ information; adaptive rounds can only add nonnegative information.

\begin{proposition}[Certified exploration from a separating query family]
\label{prop:sweep}
For every $s\in\mathcal S$,
\begin{equation}
\Prob_{\mS}(\widehat m_{i,t}\neq\mS)
\le \Kprior(\mS,s)\exp\!\left(s\Delta_i-C_td_{\mathcal Q}W(s)\right).
\label{eq:sweep-finite}
\end{equation}
If certified rounds occupy asymptotic fraction $\eta$ and $C_t\ge\eta t/L-c_0$, then
\begin{equation}
E_i\ge\frac{\eta d_{\mathcal Q}}{L}\sup_{s\in\mathcal S}W(s).
\label{eq:sweep-exp}
\end{equation}
\end{proposition}
For ordered 1D thresholds, $L=M-1$ and $d_{\mathcal Q}=1$; a balanced threshold cycle is max--min optimal within synchronized threshold exploration (Appendix~\ref{app:threshold-optimality}). For an $S\times S$ grid, the $S-1$ vertical and $S-1$ horizontal cuts form a separating family with $L=2(S-1)$ and $d_{\mathcal Q}=1$. Diagonal half-planes can still be used on adaptive rounds.

For rollouts with known ground truth we report the post-hoc worst-competitor information score
\begin{equation}
\widehat\gamma_T^{\rm diag}:=\sup_{s\in\mathcal S}\frac1T\min_{m\neq m^\star}\Info_T(m;s).
\label{eq:empirical-certificate}
\end{equation}
Because $s$ is optimized after observing the rollout, this is a \emph{ranking diagnostic}, not the fixed-$s$ anytime certificate in Theorem~\ref{thm:anytime}.

\section{Experiments}\label{sec:experiments}

We test three predictions in ordered 1D and non-ordered 2D queries:
approximately exponential localization under linear information growth,
graph-dependent finite-horizon behavior, and stronger association with
the least-resolved competitor than with aggregate information. Rates
follow the deterministic pre-floor fitting rule in
Appendix~\ref{app:experimental-details}. Uncertainty uses independent
trajectories after within-trajectory agent averaging; pooled
correlations use $10{,}000$ paired seed-cluster-bootstrap replicates.
Curve intervals are computed on the original scale, with log-axis
clipping only for display.

\subsection{1D ordered thresholds: convergence, graph effects, and the bottleneck}
\label{sec:exp1d}

\paragraph{Exponential convergence across graphs.}
We first fix exploration+bisection and vary only the communication graph
($M=32$, $N=10$, homogeneous $\epsilon=0.25$, $5{,}000$ trials, $80$
rounds). Before the Monte Carlo floor, MAP error is approximately
log-linear with fitted exponents $0.4288$ (complete), $0.4035$ (ring),
and $0.3384$ (directed); the corresponding post-hoc
hardest-competitor scores are $0.5380$, $0.5295$, and $0.5230$.
Since $\Delta_i$ is an additive finite-time offset, it does not change
the asymptotic exponent. The observed finite-window slope differences
can therefore reflect pre-asymptotic mixing effects and/or
graph-dependent lower-tail behavior of realized information, rather
than a change of asymptotic rate caused by $\Delta_i$ itself.
Figure~\ref{fig:graph-convergence} shows the curves; the squared-loss
companion is in the appendix.

\begin{figure}[t]
\centering
\includegraphics[width=0.94\columnwidth]
{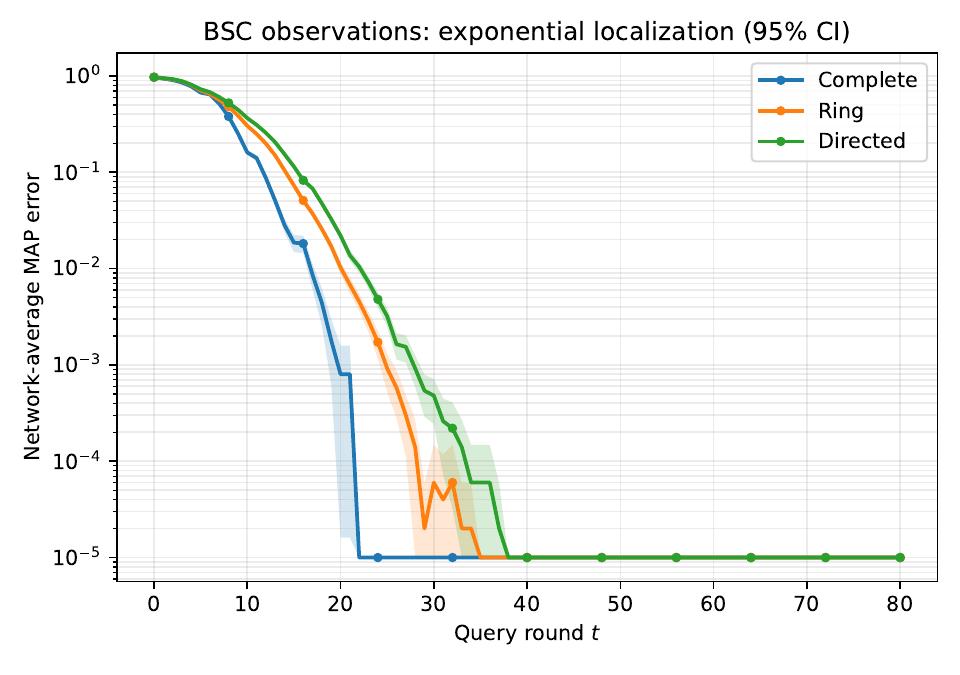}
\caption{\textbf{1D exponential localization across graph geometries.}
With the sensing policy fixed, network-average MAP error decays
approximately exponentially before the Monte Carlo floor, with
graph-dependent finite-horizon rates. Shading denotes pointwise
95\% confidence intervals across independent trajectories.}
\label{fig:graph-convergence}
\end{figure}

\begin{figure*}[t]
\centering
\includegraphics[width=0.96\textwidth]
{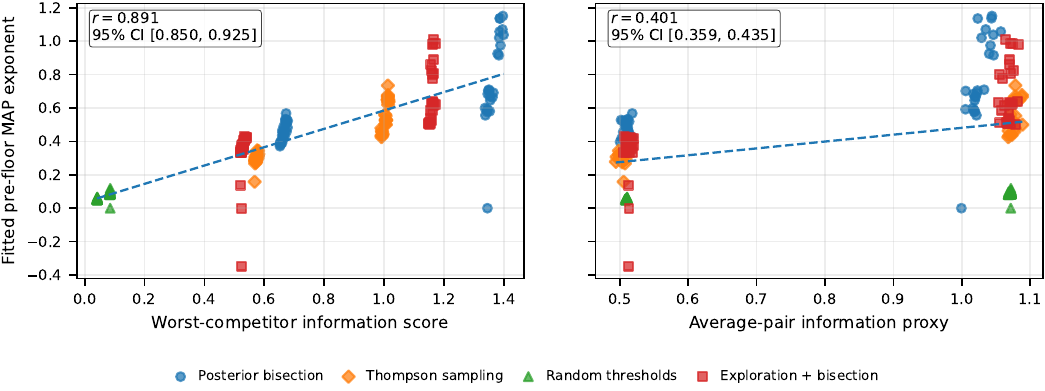}
\caption{\textbf{1D hardest-competitor bottleneck.}
Across policies, graph topologies, sensor profiles, and seed blocks,
empirical localization speed correlates more strongly with the
worst-competitor information score than with the average-pair proxy.
The paired difference is $\Delta r=0.490$ with 95\% CI
$[0.446,0.527]$.}
\label{fig:proxy-comparison}
\end{figure*}

\paragraph{Worst-competitor study.}
We next vary policy, graph, sensor profile, and seed over $240$
configurations. The deterministic fit rule returns a resolvable
pre-floor exponent for $210$; the remaining $30$ are reported as
n/a rather than assigned an imputed rate. Thus all 1D correlation
statements are explicitly conditional on this $210$-configuration
subset. On it, the post-hoc hardest-competitor score in
Eq.~\eqref{eq:empirical-certificate} has Pearson
$r_{\rm worst}=0.891$ (95\% seed-cluster-bootstrap CI
$[0.850,0.925]$) with the empirical localization exponent. As an
ablation, we average the same pairwise information over all pairs,
\begin{align}
\overline{\Info}_T(s)
&:=\binom{M}{2}^{-1}
\sum_{m<m'}\Info_T(m,m';s),\nonumber\\
\widehat\gamma_T^{\rm avg}
&:=\sup_{s\in\mathcal S}
\frac{\overline{\Info}_T(s)}{T},
\label{eq:average-pair-proxy}
\end{align}
which gives $r_{\rm avg}=0.401$ (95\% CI $[0.359,0.435]$).
The paired difference is $\Delta r=0.490$ (95\% CI
$[0.446,0.527]$), and the hardest-score regression slope is $0.549$
(95\% CI $[0.524,0.578]$).
Figure~\ref{fig:proxy-comparison} summarizes this pooled linear
association. Because points cluster by policy, these Pearson
statistics should not be read as a within-policy rank-ordering
guarantee.

\subsection{2D spatial queries: policy behavior and the bottleneck}
\label{sec:exp-2d}

Ordered thresholds have special one-dimensional structure, so we next
ask whether the same information bottleneck persists when neither the
targets nor the admissible queries are linearly ordered. The target set
is an $8\times8$ grid ($M=64$), and the adaptive library contains $42$
vertical, horizontal, and diagonal half-plane queries. The certified
family contains the $14$ vertical/horizontal cuts and separates every
pair. Posterior bisection chooses, from this finite spatial library, the
query whose one-side posterior mass is closest to $1/2$.
Unlike Fig.~\ref{fig:graph-convergence}, which isolates network effects
in 1D with the sensing policy fixed, here we first illustrate
policy-dependent localization on a fixed complete graph and then test
the information bottleneck across policies, graphs, sensors, and seeds.

\begin{figure}[t]
\centering
\includegraphics[width=0.96\columnwidth]
{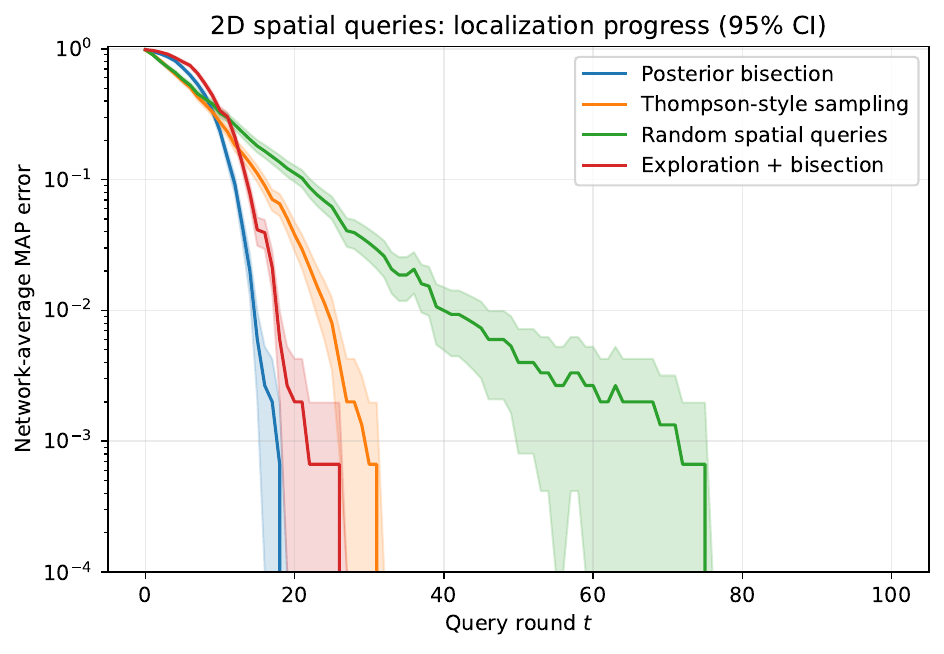}
\caption{\textbf{Policy-dependent localization in 2D.}
Network-average MAP error for posterior bisection, Thompson-style
sampling, random spatial querying, and exploration+bisection on the
$8\times8$ spatial grid with a complete communication graph. Shading
denotes pointwise 95\% confidence intervals across independent Monte
Carlo trajectories.}
\label{fig:2d-localization-map}
\end{figure}

\paragraph{Policy-dependent 2D localization.}
Figure~\ref{fig:2d-localization-map} shows network-average MAP error
for posterior bisection, Thompson-style sampling, random spatial
querying, and exploration+bisection on the complete graph. The four
policies exhibit distinct pre-floor localization behavior. Random
spatial querying is particularly diagnostic: its average-pair score is
$0.44$, whereas its hardest-competitor score and fitted exponent are
both about $0.09$. Thus aggregate pairwise coverage can appear
substantial even when one poorly resolved alternative controls the
observed localization rate.

\paragraph{2D worst-competitor study.}
We repeat the pooled design over four policies, three communication
graphs, two sensor profiles, and ten seed blocks, giving $240$
configurations. All $240$ have valid pre-floor fits. The
hardest-competitor score has $r_{\rm worst}=0.768$
(95\% CI $[0.761,0.777]$), while the same average-pair proxy gives
$r_{\rm avg}=0.484$ (95\% CI $[0.474,0.494]$). Their paired
difference is $\Delta r=0.285$ (95\% CI $[0.270,0.299]$), and the
hardest-score regression slope is $0.773$ (95\% CI
$[0.736,0.811]$). Graph effects persist in the pooled sweep, with mean
exponents $0.44$, $0.33$, and $0.28$ for complete, ring, and directed
graphs, respectively. The same bottleneck behavior therefore persists
after replacing ordered thresholds with a non-ordered spatial query
library.

\begin{figure}[t]
\centering
\includegraphics[width=0.96\columnwidth]
{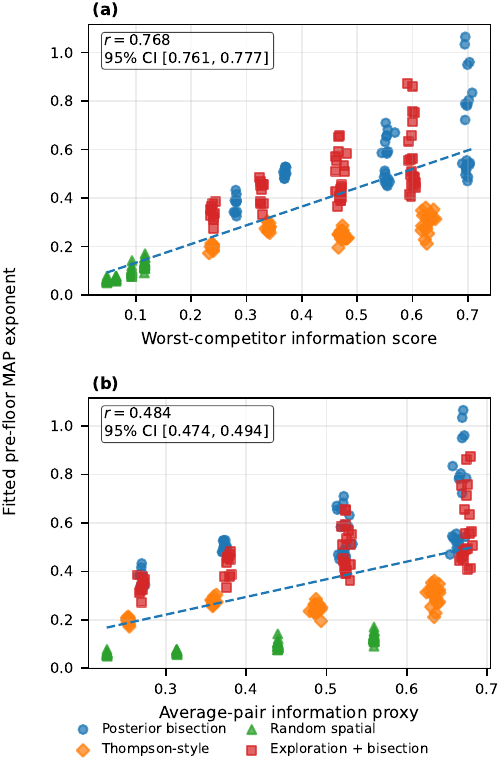}
\caption{\textbf{2D hardest-competitor bottleneck.}
Across $240$ policy--graph--sensor--seed configurations on an
$8\times8$ spatial grid, worst-competitor information correlates more
strongly with empirical localization speed than the average-pair
proxy. The paired difference is $\Delta r=0.285$ with 95\% CI
$[0.270,0.299]$.}
\label{fig:proxy-comparison-2d}
\end{figure}

\section{Limitations}\label{sec:limitations}
The theory assumes a finite hypothesis class, fixed primitive graph, common full-support prior, and known conditionally independent sensors. The fixed graph is structural because the proof uses one stationary left eigenvector; time-varying graphs need a product-mixing argument. Bounded pairwise log-likelihood ratios exclude Gaussian tails directly. Online stopping also needs global query logging, naive $O(M^2)$ pair tracking, and a pre-fixed $s$. Learned-policy exponents further require a lower-tail information-growth guarantee. Continuous hypotheses, misspecification, and these deployment extensions remain open.

\section{Conclusion}
We introduced a pathwise information certificate for decentralized adaptive sensing that evaluates the evidence actually collected by an arbitrary history-dependent policy. The same realized-information process yields fixed-horizon MAP bounds, an anytime network-wide stopping rule, and, under linear information growth, exponential localization rates. A pairwise KL converse explains why the hardest remaining alternative is the relevant bottleneck. Across the tested 1D and 2D query families, worst-competitor information has a stronger pooled association with localization speed than average-pair information, supporting the certificate as a reliability tool and finite-trajectory diagnostic.


\clearpage

\section*{Acknowledgments}
Research was sponsored by the Department of the Air Force Artificial Intelligence Accelerator and was
accomplished under Cooperative Agreement Number FA8750-19-2-1000. The views and conclusions contained in this document are those of the authors and should not be interpreted as representing the official policies, either expressed or implied, of the Department of the Air Force or the U.S. Government. The U.S. Government is authorized to reproduce and distribute reprints for Government purposes notwithstanding any copyright notation herein.

\section*{AI Use Statement}
In this work, we used generative AI tools to polish the writing (language, grammatical refinement), and debug experimental code. We did not use AI for developing research ideas, designing methodology, proving mathematical statements, conducting experiments, or analyzing results. We have reviewed all AI-assisted work to ensure accuracy and faithful representation of our claims. All mathematical and algorithmic claims, experimental findings, and paper contents were conceived and verified by the authors. We take responsibility for the final content of this work.

\bibliographystyle{apalike}
\bibliography{references_final_revised_2d}

\section*{Checklist}

\begin{enumerate}

  \item For all models and algorithms presented, check if you include:
  \begin{enumerate}
    \item A clear description of the mathematical setting, assumptions,
    algorithm, and/or model. [Yes] The decentralized adaptive querying
    model, BSC observation model, local Bayes update, and geometric
    pooling rule are fully specified in Section~\ref{sec:model}, with the Bayes/social-pooling update in Eq.~\eqref{eq:pool}; the controlled-observation extension is summarized in Section~\ref{sec:theory} and derived in Appendix~\ref{app:general-observation}.
    \item An analysis of the properties and complexity (time, space,
    sample size) of any algorithm. [Yes] Per-round computational complexity is given in Section \ref{sec:model}, and the stopping-time/sample-complexity consequence is derived after Theorem \ref{thm:anytime} in Section \ref{sec:anytime}.
    \item (Optional) Anonymized source code, with specification of all
    dependencies, including external libraries. [Yes] Anonymized code is included in the supplementary material.
  \end{enumerate}

  \item For any theoretical claim, check if you include:
  \begin{enumerate}
    \item Statements of the full set of assumptions of all theoretical
    results. [Yes] Primitive/stationary graph, common full-support
    prior, conditionally independent sensing noise, and (for the controlled-observation extension summarized in Section~\ref{sec:theory} and derived in Appendix~\ref{app:general-observation}) bounded log-likelihood ratios are all stated explicitly.
    \item Complete proofs of all theoretical results. [Yes] Full proofs
    of the finite-time and anytime certificates, the stochastic-rate and
    MAP/MSE corollaries, the aggregate-information counterexample, the
    pairwise minimax converse, and the certified-exploration results are
    given in Appendix~\ref{app:proofs}; the explicit mixing calculation is included there
    as well.
    \item Clear explanations of any assumptions. [Yes] Each theorem is
    followed by an interpretive paragraph relating assumptions to the
    decentralized adaptive-querying setting (e.g., the discussion after
    Theorem~\ref{thm:main} and Proposition~\ref{prop:sweep}).
  \end{enumerate}

  \item For all figures and tables that present empirical results,
  check if you include:
  \begin{enumerate}
    \item The code, data, and instructions needed to reproduce the main
    experimental results (either in the supplemental material or as a
    URL). [Yes] The supplementary material includes the code and reproduction instructions, with graph construction, sensor, and seed configuration specified.
    \item All the training details (e.g., data splits, hyperparameters,
    how they were chosen). [Yes] Simulation parameters (grid geometry,
    agents $N$, BSC noise $\epsilon$, query library, trial counts, round
    horizons, and seeds) are specified per experiment in Appendix~\ref{app:experimental-details}.
    \item A clear definition of the specific measure or statistics and
    error bars (e.g., with respect to the random seed after running
    experiments multiple times). [Yes] We clearly define all empirical statistics and uncertainty estimates. MAP-error and localization-loss curves report pointwise 95\% intervals across independent Monte Carlo trajectories, with agents averaged within each trajectory. The 1D code uses Student-$t$ critical values (normal fallback), while the 2D plotting code uses mean $\pm1.96$ standard errors; all intervals are computed on the original linear scale and any log-plot clipping is display-only. For both pooled ten-seed bottleneck studies, we report 95\% seed-cluster-bootstrap confidence intervals for Pearson correlation and the regression slope using $10{,}000$ bootstrap replicates; the worst-competitor versus average-pair comparison additionally reports a paired seed-cluster-bootstrap confidence interval for $\Delta r$.
    \item A description of the computing infrastructure used (e.g.,
    type of GPUs, internal cluster, or cloud provider).
    [Yes] Experiments used CPU execution on a single 4-core workstation.
  \end{enumerate}

  \item If you are using existing assets (e.g., code, data, models) or
  curating/releasing new assets, check if you include:
  \begin{enumerate}
    \item Citations of the creator if your work uses existing assets.
    [Not Applicable] The paper does not build on external datasets,
    pretrained models, or third-party code assets beyond standard
    numerical libraries.
    \item The license information of the assets, if applicable.
    [Not Applicable] No external licensed assets are used.
    \item New assets either in the supplemental material or as a URL,
    if applicable. [Yes] Simulation code implementing the query
    policies, graph constructions, and figure-generation scripts is
    released as a new asset in the supplementary material.
    \item Information about consent from data providers/curators.
    [Not Applicable] All data are synthetically generated via
    controlled simulation; no external data providers are involved.
    \item Discussion of sensible content if applicable, e.g., personally
    identifiable information or offensive content. [Not Applicable]
    The work involves no human subject data, PII, or sensitive content.
  \end{enumerate}

  \item If you used crowdsourcing or conducted research with human
  subjects, check if you include:
  \begin{enumerate}
    \item The full text of instructions given to participants and
    screenshots. [Not Applicable] No human subjects or crowdsourcing
    were involved in this work.
    \item Descriptions of potential participant risks, with links to
    Institutional Review Board (IRB) approvals if applicable.
    [Not Applicable] No human subjects research was conducted.
    \item The estimated hourly wage paid to participants and the total
    amount spent on participant compensation. [Not Applicable] No
    participant compensation was involved.
  \end{enumerate}

\end{enumerate}

\clearpage
\appendix
\thispagestyle{empty}
\onecolumn
\aistatstitle{Pathwise Information Certificates for Decentralized Adaptive Sensing: \\
Supplementary Materials}

\section{Complete Technical Proofs}
\label{app:proofs}

\subsection{Posterior-odds recursion with a common full-support prior}
Fix a false hypothesis $m\neq\mS$ and define the posterior odds and log-odds
\[
R_{i,t}(m):=\frac{\mu_{i,t}(m)}{\mu_{i,t}(\mS)},
\qquad
r_{i,t}(m):=\log R_{i,t}(m).
\]
For agent $j$, define the one-step likelihood ratio
\[
L_{j,t}(m)
:=\frac{\ell_{j,t}(Y_{j,t+1}\mid m)}
        {\ell_{j,t}(Y_{j,t+1}\mid\mS)}.
\]
We first derive the recursion used throughout the proofs. From the local Bayes update,
\[
\frac{\widetilde\mu_{j,t+1}(m)}
     {\widetilde\mu_{j,t+1}(\mS)}
=
\frac{\mu_{j,t}(m)\ell_{j,t}(Y_{j,t+1}\mid m)}
     {\mu_{j,t}(\mS)\ell_{j,t}(Y_{j,t+1}\mid\mS)}
=R_{j,t}(m)L_{j,t}(m),
\]
because the local normalizing denominator cancels. Taking the ratio of the two pooled beliefs in \eqref{eq:pool} similarly cancels the social-pooling normalizer, giving
\[
R_{i,t+1}(m)
=
\prod_{j=1}^{N}
\left[R_{j,t}(m)L_{j,t}(m)\right]^{A_{ij}}.
\]
Taking logarithms and defining
$\rb_t(m)=(r_{1,t}(m),\ldots,r_{N,t}(m))^\top$ and
$\boldsymbol\lambda_t(m)=(\log L_{1,t}(m),\ldots,\log L_{N,t}(m))^\top$ yields
\begin{equation}
\rb_{t+1}(m)=A\big(\rb_t(m)+\boldsymbol\lambda_t(m)\big).
\label{eq:ratio-app}
\end{equation}
Because all agents use the same prior,
\[
\rb_0(m)
=
\log\!\frac{\prior(m)}{\prior(\mS)}\,\mathbf1.
\]

\subsection{Fractional-moment martingale}
Define the stationary geometric posterior odds
\[
G_t(m)
:=\prod_{i=1}^{N}R_{i,t}(m)^{v_i}
=\exp\!\big(\vpi^\top\rb_t(m)\big).
\]
Premultiplying \eqref{eq:ratio-app} by $\vpi^\top$ and using $\vpi^\top A=\vpi^\top$ gives
\[
\vpi^\top\rb_{t+1}(m)
=
\vpi^\top\rb_t(m)+\sum_{j=1}^{N}v_j\log L_{j,t}(m),
\]
and therefore
\begin{equation}
G_{t+1}(m)
=G_t(m)\prod_{j=1}^{N}L_{j,t}(m)^{v_j}.
\label{eq:G-rec}
\end{equation}

Fix $s\in\mathcal S$, so $\alpha_j:=sv_j\in[0,1]$ for every agent. Conditional on $\cF_t$, every query selected at round $t$ is fixed. If $q_{j,t}(m)=q_{j,t}(\mS)$, the two hypotheses induce the same BSC output law, hence $L_{j,t}(m)=1$ almost surely and
$\E_{\mS}[L_{j,t}(m)^{\alpha_j}\mid\cF_t]=1$.

If the query separates the pair, then under the true hypothesis $\mS$ the likelihood ratio equals
\[
L_{j,t}(m)=
\begin{cases}
\eps_j/(1-\eps_j), & \text{with probability }1-\eps_j,\\
(1-\eps_j)/\eps_j, & \text{with probability }\eps_j.
\end{cases}
\]
Consequently,
\[
\E_{\mS}[L_{j,t}(m)^{\alpha_j}\mid\cF_t]
=(1-\eps_j)^{1-\alpha_j}\eps_j^{\alpha_j}
+\eps_j^{1-\alpha_j}(1-\eps_j)^{\alpha_j}
=e^{-\J_j(\alpha_j)}.
\]
Combining the separating and nonseparating cases gives
\[
\E_{\mS}[L_{j,t}(m)^{sv_j}\mid\cF_t]
=
\exp\!\left[-\J_j(sv_j)I_{j,t}(m,\mS)\right].
\]
The BSC noises are conditionally independent across agents, so raising \eqref{eq:G-rec} to the power $s$ and taking conditional expectation yields
\[
\E_{\mS}[G_{t+1}(m)^s\mid\cF_t]
=
G_t(m)^s
\exp\!\left[-\sum_{j=1}^{N}\J_j(sv_j)I_{j,t}(m,\mS)\right].
\]
Because
\[
\Info_{t+1}(m;s)
=
\Info_t(m;s)
+\sum_{j=1}^{N}\J_j(sv_j)I_{j,t}(m,\mS),
\]
the process
\begin{equation}
Z_t(m;s):=G_t(m)^s\exp\!\big(\Info_t(m;s)\big)
\label{eq:martingale}
\end{equation}
satisfies
\[
\E_{\mS}[Z_{t+1}(m;s)\mid\cF_t]=Z_t(m;s).
\]
Thus $\{Z_t(m;s)\}_{t\ge0}$ is a nonnegative martingale. At $t=0$, since $\sum_i v_i=1$,
\[
G_0(m)
=\frac{\prior(m)}{\prior(\mS)},
\]
so for every $t$,
\begin{equation}
\E_{\mS}Z_t(m;s)=Z_0(m;s)
=\left(\frac{\prior(m)}{\prior(\mS)}\right)^s.
\label{eq:martingale-prior}
\end{equation}

\subsection{Network disagreement is a bounded transient}
Unrolling \eqref{eq:ratio-app} gives
\[
\rb_t(m)
=A^t\rb_0(m)
+\sum_{\tau=0}^{t-1}A^{t-\tau}\boldsymbol\lambda_\tau(m).
\]
The common-prior vector is proportional to $\mathbf1$. Since $A^t\mathbf1=\mathbf1$ and $\vpi^\top\mathbf1=1$, its contribution is identical in $r_{i,t}(m)$ and $\vpi^\top\rb_t(m)$ and therefore cancels. Hence
\[
r_{i,t}(m)-\vpi^\top\rb_t(m)
=
\sum_{\tau=0}^{t-1}\sum_{j=1}^{N}
\big((A^{t-\tau})_{ij}-v_j\big)\lambda_{j,\tau}(m).
\]
For the BSC model,
\[
|\lambda_{j,\tau}(m)|
=|\log L_{j,\tau}(m)|
\le \log\frac{1-\eps_j}{\eps_j}=b_j.
\]
Therefore, by the triangle inequality,
\[
\left|r_{i,t}(m)-\vpi^\top\rb_t(m)\right|
\le
\sum_{\tau=0}^{t-1}\sum_{j=1}^{N}
\left|(A^{t-\tau})_{ij}-v_j\right|b_j.
\]
Letting $k=t-\tau$ and enlarging the finite sum to the infinite series in \eqref{eq:Delta},
\begin{equation}
\left|r_{i,t}(m)-\vpi^\top\rb_t(m)\right|\le\Delta_i.
\label{eq:network-bound}
\end{equation}
Because $\vpi^\top\rb_t(m)=\log G_t(m)$, the upper side of \eqref{eq:network-bound} implies
\[
r_{i,t}(m)\le\log G_t(m)+\Delta_i.
\]
Exponentiating and raising to $s>0$ gives
\begin{equation}
R_{i,t}(m)^s\le e^{s\Delta_i}G_t(m)^s.
\label{eq:ratio-to-G}
\end{equation}

\subsection{Proof of Theorem~\ref{thm:main}}
Fix $i$, $t$, and the true hypothesis $\mS$. If the MAP estimate is incorrect, then at least one false hypothesis has posterior mass at least that of the truth. Therefore
\[
\{\widehat m_{i,t}\neq\mS\}
\subseteq
\bigcup_{m\neq\mS}\{R_{i,t}(m)\ge1\},
\]
and the union bound gives
\begin{equation}
\Prob_{\mS}(\widehat m_{i,t}\neq\mS)
\le
\sum_{m\neq\mS}\Prob_{\mS}(R_{i,t}(m)\ge1).
\label{eq:union-app}
\end{equation}
Now fix one false $m$. Split the event according to whether enough information has accumulated:
\[
\Prob_{\mS}(R_{i,t}(m)\ge1)
\le
\Prob_{\mS}(\Info_t(m;s)<a_{m,t})
+
\Prob_{\mS}(R_{i,t}(m)\ge1,\ \Info_t(m;s)\ge a_{m,t}).
\]
On the second event, $R_{i,t}(m)^s\ge1$, so
\[
\Prob_{\mS}(R_{i,t}(m)\ge1,\ \Info_t(m;s)\ge a_{m,t})
\le
\E_{\mS}\!\left[
R_{i,t}(m)^s
\1\{\Info_t(m;s)\ge a_{m,t}\}
\right].
\]
Using \eqref{eq:ratio-to-G},
\[
\E_{\mS}\!\left[
R_{i,t}(m)^s\1\{\Info_t\ge a_{m,t}\}
\right]
\le
e^{s\Delta_i}
\E_{\mS}\!\left[
G_t(m)^s\1\{\Info_t\ge a_{m,t}\}
\right].
\]
From \eqref{eq:martingale}, $G_t(m)^s=e^{-\Info_t(m;s)}Z_t(m;s)$. Hence
\[
\begin{aligned}
&e^{s\Delta_i}
\E_{\mS}\!\left[G_t(m)^s\1\{\Info_t\ge a_{m,t}\}\right]\\
&\qquad=
e^{s\Delta_i}
\E_{\mS}\!\left[e^{-\Info_t(m;s)}Z_t(m;s)
\1\{\Info_t\ge a_{m,t}\}\right]\\
&\qquad\le
e^{s\Delta_i-a_{m,t}}\E_{\mS} Z_t(m;s).
\end{aligned}
\]
Applying \eqref{eq:martingale-prior} gives
\[
\Prob_{\mS}(R_{i,t}(m)\ge1)
\le
\Prob_{\mS}(\Info_t(m;s)<a_{m,t})
+
\left(\frac{\prior(m)}{\prior(\mS)}\right)^s
 e^{s\Delta_i-a_{m,t}}.
\]
Summing this inequality over $m\neq\mS$ in \eqref{eq:union-app} proves \eqref{eq:main-bound}. \qed

\subsection{Proof of the anytime-valid network-wide certificate}
\label{app:anytime-proof}
We first prove a pairwise time-uniform statement. Fix agent $i$, false hypothesis $m\neq\mS$, and a constant $c\ge0$. If at some time $t$ both
$R_{i,t}(m)\ge1$ and $\Info_t(m;s)\ge c$, then \eqref{eq:ratio-to-G} implies
\[
1\le R_{i,t}(m)^s
\le e^{s\Delta_i}G_t(m)^s.
\]
Therefore $G_t(m)^s\ge e^{-s\Delta_i}$. Multiplying by $e^{\Info_t(m;s)}$ shows
\[
Z_t(m;s)
=G_t(m)^s e^{\Info_t(m;s)}
\ge e^{c-s\Delta_i}.
\]
Hence the event
\[
\{\exists t\ge0:\ R_{i,t}(m)\ge1,\ \Info_t(m;s)\ge c\}
\]
is contained in
\[
\left\{\sup_{t\ge0}Z_t(m;s)\ge e^{c-s\Delta_i}\right\}.
\]
Since $Z_t(m;s)$ is a nonnegative martingale, Ville's inequality \citep{ville1939,howard2020timeuniform} gives
\[
\Prob_{\mS}\!\left(\sup_{t\ge0}Z_t(m;s)\ge a\right)
\le \frac{\E_{\mS} Z_0(m;s)}{a}
\qquad(a>0).
\]
Taking $a=e^{c-s\Delta_i}$ and using \eqref{eq:martingale-prior},
\begin{equation}
\Prob_{\mS}\!\left(
\exists t\ge0:\ R_{i,t}(m)\ge1,\ \Info_t(m;s)\ge c
\right)
\le
\left(\frac{\prior(m)}{\prior(\mS)}\right)^s
 e^{s\Delta_i-c}.
\label{eq:anytime-pairwise}
\end{equation}

We next pass from one false hypothesis and one agent to all of them. If agent $i$ makes an incorrect MAP decision at time $t$, then some false $m$ satisfies $R_{i,t}(m)\ge1$. Moreover, by definition of \eqref{eq:pairwise-information},
\[
\underline{\Info}_t(s)\ge c
\quad\Longrightarrow\quad
\Info_t(m,\mS;s)\ge c
\quad\text{for every }m\neq\mS.
\]
The true-vs-false pairwise quantity $\Info_t(m,\mS;s)$ is exactly the $\Info_t(m;s)$ used in the martingale proof. Consequently, a union bound over agents and false hypotheses, followed by \eqref{eq:anytime-pairwise}, yields
\[
\begin{aligned}
&\Prob_{\mS}\!\left(
\exists i,\exists t:\widehat m_{i,t}\neq\mS,
\ \underline{\Info}_t(s)\ge c
\right)\\
&\qquad\le
\sum_{i=1}^{N}\sum_{m\neq\mS}
\left(\frac{\prior(m)}{\prior(\mS)}\right)^s
 e^{s\Delta_i-c}\\
&\qquad\le
N\mathcal{K}_{\max}(s)e^{s\Delta_{\max}-c}.
\end{aligned}
\]
Set $c=c_\delta(s)$ from \eqref{eq:anytime-threshold}. The final expression equals $\delta$.

It remains only to connect this event to the stopping time. Every term in \eqref{eq:pairwise-information} is nonnegative, so $\underline{\Info}_t(s)$ is nondecreasing in $t$. Therefore, on $\{\tau_\delta(s)<\infty\}$,
\[
t\ge\tau_\delta(s)
\quad\Longrightarrow\quad
\underline{\Info}_t(s)\ge c_\delta(s).
\]
Thus
\[
\{\exists i,\exists t\ge\tau_\delta(s):\widehat m_{i,t}\neq\mS\}
\subseteq
\{\exists i,\exists t:\widehat m_{i,t}\neq\mS,
\ \underline{\Info}_t(s)\ge c_\delta(s)\},
\]
whose probability is at most $\delta$. This proves Theorem~\ref{thm:anytime}. \qed

\subsection{Derivation of the anytime sample-complexity consequence}
Assume $\underline{\Info}_t(s)\ge\gamma t-B$ for every $t$. By the definition of $\tau_\delta(s)$ in \eqref{eq:anytime-threshold}, it is enough to find an integer $t$ for which
\[
\gamma t-B\ge c_\delta(s).
\]
Indeed, the assumed information-growth inequality then implies
\[
\underline{\Info}_t(s)
\ge \gamma t-B
\ge c_\delta(s),
\]
so the stopping condition has been met no later than time $t$. Solving the first inequality gives
\[
t\ge
\frac{B+c_\delta(s)}{\gamma}.
\]
Substituting the definition of $c_\delta(s)$ from \eqref{eq:anytime-threshold},
\[
t\ge
\frac{B+s\Delta_{\max}
+\log(N\mathcal K_{\max}(s)/\delta)}
{\gamma}.
\]
Taking the smallest integer satisfying this inequality yields
\[
\tau_\delta(s)
\le
\left\lceil
\frac{B+s\Delta_{\max}
+\log(N\mathcal K_{\max}(s)/\delta)}
{\gamma}
\right\rceil,
\]
which is the sample-complexity statement given after Theorem~\ref{thm:anytime}. \qed

\subsection{Explicit mixing bound}\label{app:mixing-proof}
For a finite primitive stochastic matrix, Perron--Frobenius theory gives a simple eigenvalue at $1$ and places every other eigenvalue strictly inside the unit disk \citep{seneta2006nonnegative}. Consequently, the powers converge geometrically to the stationary projection: there exist $C_A<\infty$ and $\rho_A<1$ such that
\[
\max_i\left\|e_i^\top A^k-\vpi^\top\right\|_1\le C_A\rho_A^k.
\]
If a nonsymmetric $A$ has nontrivial Jordan blocks, the spectral expansion can contain a polynomial factor $k^{r-1}\lambda^k$; for any $\rho_A$ strictly larger than the largest nonunit eigenvalue modulus, that polynomial factor is absorbed into a finite constant $C_A$. Thus the displayed geometric bound still holds. This is the standard finite-state geometric-ergodicity property of an irreducible aperiodic chain \citep{levin2017markov}.

By definition of $\Delta_i$ and $b_j\le b_{\max}$,
\[
\Delta_i
\le
b_{\max}\sum_{k=1}^{\infty}
\left\|e_i^\top A^k-\vpi^\top\right\|_1
\le
b_{\max}C_A\sum_{k=1}^{\infty}\rho_A^k
=
b_{\max}C_A\frac{\rho_A}{1-\rho_A},
\]
which proves the stated mixing bound.

Now suppose $A$ is symmetric and doubly stochastic. Then $\vpi=\mathbf1/N$. Define $\lambda_\star:=\max_{\ell\ge2}|\lambda_\ell(A)|<1$, the largest absolute nonunit eigenvalue. The component orthogonal to $\mathbf1$ contracts in Euclidean norm by at most $\lambda_\star^k$. Hence
\[
\left\|e_i^\top A^k-\vpi^\top\right\|_2
\le
\lambda_\star^k
\left\|e_i-\frac{\mathbf1}{N}\right\|_2
=
\lambda_\star^k\sqrt{1-\frac1N}.
\]
Using $\|x\|_1\le\sqrt N\|x\|_2$,
\[
\left\|e_i^\top A^k-\vpi^\top\right\|_1
\le\sqrt{N-1}\,\lambda_\star^k.
\]
Summing the geometric series gives
\[
\Delta_i
\le
b_{\max}\sqrt{N-1}\frac{\lambda_\star}{1-\lambda_\star}.
\]
\qed

\subsection{Proof of Corollary~\ref{cor:stochastic-growth}}
For all sufficiently large $t$ such that $\gamma t-B\ge0$, set
$a_{m,t}=\gamma t-B$ for every false $m$ in Theorem~\ref{thm:main}. Then
\[
\begin{aligned}
\Prob_{\mS}(\widehat m_{i,t}\neq\mS)
&\le
\sum_{m\neq\mS} C_m e^{-\kappa t}
+\sum_{m\neq\mS}
\left(\frac{\prior(m)}{\prior(\mS)}\right)^s
 e^{s\Delta_i-(\gamma t-B)}\\
&=
\left(\sum_{m\neq\mS}C_m\right)e^{-\kappa t}
+\Kprior(\mS,s)e^{s\Delta_i+B}e^{-\gamma t}.
\end{aligned}
\]
This finite-$t$ bound immediately yields the rate claim. If $c_1,c_2>0$, then
$c_1e^{-\kappa t}+c_2e^{-\gamma t}$ has asymptotic exponential rate
$\min\{\kappa,\gamma\}$, because the slower-decaying exponential dominates the sum up to multiplicative constants. Therefore
$E_i\ge\min\{\gamma,\kappa\}$. If the information lower bound holds deterministically, the lower-tail probabilities are zero and the $e^{-\kappa t}$ term disappears. \qed

\subsection{Proof of Corollary~\ref{cor:mse}}
On the event $\{\widehat m_{i,t}\neq\mS\}$, the estimated and true hypotheses are distinct, so by definition of the minimum separation in the chosen metric,
\[
d(\widehat\theta_{i,t},\theta_{\mS})\ge d_{\min}.
\]
For every outcome, both points belong to the finite target set, so
\[
d(\widehat\theta_{i,t},\theta_{\mS})\le D_\Theta.
\]
Combining the two observations pointwise,
\[
d_{\min}^2\1\{\widehat m_{i,t}\neq\mS\}
\le
d(\widehat\theta_{i,t},\theta_{\mS})^2
\le
D_\Theta^2\1\{\widehat m_{i,t}\neq\mS\}.
\]
Taking expectations proves \eqref{eq:mse-sandwich}. Multiplying the finite-time MAP upper bound in Theorem~\ref{thm:main} by $D_\Theta^2$ gives the stated squared-metric upper bound. Under the stated assumption $d_{\min}>0$, both sandwich constants are fixed and positive, so multiplication by them does not change $\liminf_{t\to\infty}-t^{-1}\log(\cdot)$. Hence MAP error and squared-metric localization loss have the same exponent. If $d_{\min}=0$, only the upper-bound implication is asserted. \qed

\subsection{Proof of Proposition~\ref{prop:aggregate-insufficient}}
\label{app:aggregate-proof}
Let $k_\star=\lfloor M/2\rfloor$ and let every agent repeat the single threshold $q_{k_\star}$. Any two hypotheses on the same side of the cut generate the same noiseless query bit at every round and therefore induce exactly the same complete transcript distribution. Because $M\ge4$, at least one side contains two distinct hypotheses $m,m'$. If $P$ denotes their common transcript law, the disjoint events $\{\widehat m_T=m\}$ and $\{\widehat m_T=m'\}$ satisfy
\[
P(\widehat m_T=m)+P(\widehat m_T=m')\le1.
\]
Hence their two error probabilities sum to at least one, so one is at least $1/2$, proving the minimax lower bound. The cut separates exactly $k_\star(M-k_\star)$ unordered pairs, which gives Eq.~\eqref{eq:single-cut-fraction}.

Now cycle once through $q_1,\ldots,q_{M-1}$. For threshold $k$, exactly $k(M-k)$ unordered pairs lie on opposite sides. Therefore
\[
\sum_{k=1}^{M-1}k(M-k)=\frac{M(M-1)(M+1)}{6},
\]
and dividing by the $M-1$ rounds in a cycle and by $\binom{M}{2}$ gives
\[
\phi_{\rm cyc}(M)=\frac{M+1}{3(M-1)}.
\]
Every adjacent pair $(k,k+1)$ is separated by exactly one threshold in a complete cycle, so the worst-pair frequency is $1/(M-1)$; nonadjacent pairs are separated more often.

It remains to compare the average pair fractions. If $M$ is even, $k_\star=M/2$ and
\[
\phi_{\rm one}(M)=\frac{M}{2(M-1)}
>\frac{M+1}{3(M-1)}=\phi_{\rm cyc}(M)
\]
for $M>2$. If $M$ is odd, $k_\star=(M-1)/2$ and
\[
\phi_{\rm one}(M)=\frac{M+1}{2M}
>\frac{M+1}{3(M-1)}=\phi_{\rm cyc}(M)
\]
for $M>3$. Thus the strict inequality holds for every $M\ge4$. The balanced cyclic schedule nevertheless has positive worst-pair coverage, so Proposition~\ref{prop:sweep} gives a positive certified exponent when the cycle is repeated. \qed

\subsection{Proof of Proposition~\ref{prop:pairwise-converse}}
\label{app:pairwise-converse-proof}
Fix two distinct hypotheses $m,m'$ and let $\mathsf P_m^T$ and $\mathsf P_{m'}^T$ denote the laws of the complete transcript through round $T$. For any estimator $\widehat m_T$, define $A:=\{\widehat m_T=m\}$. Then
\[
\Prob_m(\widehat m_T\neq m)+\Prob_{m'}(\widehat m_T\neq m')
\ge
\mathsf P_m^T(A^c)+\mathsf P_{m'}^T(A).
\]
The Bretagnolle--Huber inequality \citep{bretagnolle1979estimation} gives
\[
\mathsf P_m^T(A^c)+\mathsf P_{m'}^T(A)
\ge
\frac12\exp\!\left[-D_{\rm KL}(\mathsf P_m^T\|\mathsf P_{m'}^T)\right].
\]
Therefore at least one of the two hypothesis-specific error probabilities is at least one quarter of the exponential term:
\begin{equation}
\max\!\left\{
\Prob_m(\widehat m_T\neq m),
\Prob_{m'}(\widehat m_T\neq m')
\right\}
\ge
\frac14 e^{-D_{\rm KL}(\mathsf P_m^T\|\mathsf P_{m'}^T)}.
\label{eq:bh-pair}
\end{equation}

It remains to identify the transcript divergence under an adaptive policy. Let $p_{j,t}(\cdot\mid m,\cF_t)$ denote agent $j$'s conditional observation law under hypothesis $m$ after the history-dependent action at round $t$ has been selected. Apply the KL chain rule sequentially to policy randomization, selected actions, observations, and communicated messages. Conditional on a realized history, the policy uses the same action-selection kernel under $m$ and $m'$, so the action-selection term contributes zero KL divergence. Likewise, communicated messages and belief updates are generated by hypothesis-independent kernels once the realized history is fixed. The only nonzero contribution therefore comes from the observation kernels. Conditional independence across agents gives
\begin{equation}
D_{\rm KL}(\mathsf P_m^T\|\mathsf P_{m'}^T)
=
\E_m\!\left[
\sum_{t=0}^{T-1}\sum_{j=1}^{N}
D_{\rm KL}\!\left(
 p_{j,t}(\cdot\mid m,\cF_t)
 \middle\|
 p_{j,t}(\cdot\mid m',\cF_t)
\right)
\right].
\label{eq:adaptive-kl-chain}
\end{equation}

For the BSC query model, a nonseparating query induces identical output laws and hence zero KL divergence. A separating query compares $\mathrm{Bernoulli}(1-\eps_j)$ with $\mathrm{Bernoulli}(\eps_j)$ up to label exchange, giving
\[
D_{\rm KL}
=(1-\eps_j)\log\frac{1-\eps_j}{\eps_j}
+\eps_j\log\frac{\eps_j}{1-\eps_j}
=(1-2\eps_j)\log\frac{1-\eps_j}{\eps_j}
=:d_j.
\]
This proves \eqref{eq:bsc-kl-converse}. Finally, since \eqref{eq:bh-pair} holds for every pair, choose the pair minimizing the transcript KL divergence to obtain \eqref{eq:pairwise-converse}. If this minimum is $o(T)$, dividing the negative log of the lower bound by $T$ shows that the uniform error probability cannot have a positive exponential decay rate. \qed

\subsection{Proof of Proposition~\ref{prop:sweep}}
Fix two distinct hypotheses $m,m'$. By definition of $d_{\mathcal Q}$ in \eqref{eq:cert-distance}, one complete cycle through $\mathcal Q_{\rm cert}$ contains at least $d_{\mathcal Q}$ queries that separate this pair. On every synchronized certified round using such a query, all agents separate the pair, so the pairwise information increases by
\[
\sum_{j=1}^{N}\J_j(sv_j)=W(s).
\]
After $C_t$ complete cycles,
\[
\Info_t(m,m';s)\ge C_td_{\mathcal Q}W(s).
\]
In particular, for the true target $\mS$ and every false $m$,
\[
\Info_t(m;s)\ge C_td_{\mathcal Q}W(s).
\]
All noncertified rounds contribute additional nonnegative information, regardless of the adaptive base policy, so discarding them preserves the lower bound.

Set $a_{m,t}=C_td_{\mathcal Q}W(s)$ in Theorem~\ref{thm:main}. The information-shortfall probability is then zero, and
\[
\Prob_{\mS}(\widehat m_{i,t}\neq\mS)
\le
\Kprior(\mS,s)
\exp\!\left(s\Delta_i-C_td_{\mathcal Q}W(s)\right),
\]
which is \eqref{eq:sweep-finite}. Corollary~\ref{cor:mse} gives the corresponding squared-metric localization-loss bound.

If $C_t\ge\eta t/L-c_0$, then
\[
\Prob_{\mS}(\widehat m_{i,t}\neq\mS)
\le
\Kprior(\mS,s)
\exp\!\left(s\Delta_i+c_0d_{\mathcal Q}W(s)\right)
\exp\!\left(-\frac{\eta d_{\mathcal Q}W(s)}{L}t\right).
\]
The first factor is independent of $t$, so
\[
E_i\ge\frac{\eta d_{\mathcal Q}W(s)}{L}.
\]
Optimizing over $s\in\mathcal S$ proves \eqref{eq:sweep-exp}; equality of the MAP and squared-metric localization-loss exponents follows from Corollary~\ref{cor:mse}. \qed

\subsection{Max--min threshold exploration}\label{app:threshold-optimality}
\begin{proposition}[Max--min threshold exploration]
\label{prop:threshold-optimality}
For $H$ synchronized certified ordered-threshold rounds with threshold counts $n_k(H)$,
\[
\min_k n_k(H)W(s)\le \frac{H}{M-1}W(s),
\]
and a balanced cycle attains this rate asymptotically.
\end{proposition}

\paragraph{Proof.}
Fix $H$ synchronized certified threshold rounds. For each boundary $k\in\{1,\ldots,M-1\}$, let $n_k(H)$ denote how many of those rounds use threshold $q_k$. Exactly one threshold separates the adjacent pair $(k,k+1)$: if the cut lies below $k$, both hypotheses are on the same side; if it lies above $k$, they are again on the same side; and the cut $k$ places $k$ and $k+1$ on opposite sides. Therefore the certified information supplied to adjacent pair $(k,k+1)$ is exactly
\[
n_k(H)W(s).
\]
Every certified round uses exactly one threshold, so
\[
\sum_{k=1}^{M-1}n_k(H)=H.
\]
The minimum of $M-1$ nonnegative numbers cannot exceed their average. Hence
\[
\min_k n_k(H)
\le
\frac{1}{M-1}\sum_{k=1}^{M-1}n_k(H)
=\frac{H}{M-1}.
\]
Multiplying by $W(s)$ proves the claimed upper bound.

Now consider a balanced cyclic schedule. After $H$ rounds, each threshold has been used either
$\lfloor H/(M-1)\rfloor$ or $\lceil H/(M-1)\rceil$ times. Thus
\[
\frac1H\min_k n_k(H)W(s)
\longrightarrow
\frac{W(s)}{M-1}
\qquad\text{as }H\to\infty.
\]
This matches the upper bound, proving max--min optimality of the exploration-only certified rate within the synchronized ordered-threshold class. The result makes no claim about the extra information generated on adaptive noncertified rounds. \qed

\section{Query Coding and Separating Families}\label{app:query-coding}
The main reliability theory permits any binary subset query $q:[M]\to\{0,1\}$, and Proposition~\ref{prop:sweep} expresses certified exploration through an arbitrary separating family. This appendix gives the corresponding coding interpretation and clarifies how the 1D and 2D schedules relate to unrestricted binary queries.

\begin{lemma}[Shortest unrestricted separating family]\label{lem:query-code}
A family of $L$ unrestricted binary queries can separate all $M$ hypotheses only if
$L\ge\lceil\log_2 M\rceil$. This lower bound is achievable.
\end{lemma}

\paragraph{Proof.}
Associate each hypothesis $m$ with the binary response vector
\[
c(m):=(q_1(m),\ldots,q_L(m))\in\{0,1\}^L.
\]
Two hypotheses are separated by at least one query if and only if their response vectors differ. Hence a separating family requires $M$ distinct vectors in a set containing only $2^L$ vectors. Therefore $M\le2^L$ and
$L\ge\lceil\log_2 M\rceil$.

Conversely, let $L=\lceil\log_2 M\rceil$ and assign each hypothesis a distinct binary string of length $L$. For coordinate $\ell$, define $q_\ell(m)$ to be the $\ell$th bit of the string assigned to $m$. Distinct hypotheses differ in at least one coordinate, so this family separates every pair. \qed

The minimum Hamming distance of the response vectors is exactly $d_{\mathcal Q}$ in \eqref{eq:cert-distance}, so Proposition~\ref{prop:sweep} can be read as a response-code guarantee: longer minimum code distance gives more certified pairwise information per cycle. The ordered 1D threshold code has length $M-1$ and minimum distance one. For the $S\times S$ spatial experiment, the axial code formed by $S-1$ vertical and $S-1$ horizontal cuts has length $2(S-1)$ and minimum distance one. These concrete geometric libraries are not cardinality-optimal among unrestricted subset queries; they are chosen because they correspond to interpretable sensing actions.

\section{Stationary-Imbalance Comparison}
\label{app:imbalance-proof}
Define $W_{\rm std}^*=\sup_{s:\,sv_j\le1}\sum_j\J_j(sv_j)$. Under inverse-stationary innovation weights $\beta_j=1/v_j$ \citep{huang2025price}, define $W_{\rm corr}^*=\sup_{0<s\le1}\sum_j\J_j(s)$. Then $W_{\rm std}^*\le W_{\rm corr}^*=\sum_j\J_j(1/2)$, with equality if and only if $v_j=1/N$ for all $j$. This compares the leading certified per-cycle information; the correction may still enlarge the finite-time mixing prefactor.

\paragraph{Proof.}
For $0<\eps_j<1/2$, the BSC Chernoff function $\J_j(\alpha)$ is symmetric around $1/2$ and uniquely maximized at $\alpha=1/2$. Therefore, for every admissible $s$,
\begin{equation}
    \sum_j\J_j(sv_j)\le\sum_j\J_j(1/2),
\end{equation}
so $W_{\rm std}^*\le\sum_j\J_j(1/2)$. Under $\beta_j=1/v_j$, the admissibility condition becomes $0<s\le1$ and each argument $sv_j\beta_j=s$, hence
\begin{equation}
    W_{\rm corr}^*=\sup_{0<s\le1}\sum_j\J_j(s)
    =\sum_j\J_j(1/2).
\end{equation}
Equality for the standard rule requires $sv_j=1/2$ for every $j$, which is possible exactly when all $v_j$ are equal; since they sum to one, $v_j=1/N$. \qed

\section{General Controlled Observation Models: Derivation}
\label{app:general-observation}
This appendix gives the controlled-observation derivation summarized at the end of Section~\ref{sec:theory}. The BSC structure makes the certificate especially interpretable, but the martingale proof is more general. Suppose query/action $q_{j,t}$ induces, conditional on $\cF_t$, an observation density or mass function $p_{j,t}(y\mid m,\cF_t)$ under hypothesis $m$, dominated by a common measure; as in the main text, we suppress $\cF_t$ in formulas when unambiguous. Define
\[
L_{j,t}(m)=\frac{p_{j,t}(Y_{j,t+1}\mid m)}{p_{j,t}(Y_{j,t+1}\mid\mS)}.
\]
and, for $\alpha\in[0,1]$,
\begin{equation}
\J_{j,t}^{m,\mS}(\alpha)
=-\log\int
p_{j,t}(y\mid\mS)^{1-\alpha}
 p_{j,t}(y\mid m)^\alpha\,d\nu(y).
\label{eq:general-J}
\end{equation}
Conditional on $\cF_t$,
\[
\E_{\mS}[L_{j,t}(m)^\alpha\mid\cF_t]
=e^{-\J_{j,t}^{m,\mS}(\alpha)}.
\]
Hence the same fractional-moment argument applies with
\[
\Info_t^{\rm gen}(m;s)
=\sum_{\tau=0}^{t-1}\sum_{j=1}^{N}
\J_{j,\tau}^{m,\mS}(sv_j).
\]
If the truth-indexed log-likelihood ratios satisfy the uniform bound
$|\log L_{j,t}(m)|\le b_j$ almost surely for all relevant $m,t$, then the same $\Delta_i$ construction applies and Theorem~\ref{thm:main}, the explicit mixing bound, Corollary~\ref{cor:stochastic-growth}, and the finite-time squared-localization argument hold verbatim after replacing $\Info_t$ by $\Info_t^{\rm gen}$.

For the anytime result, fix any ordered pair $m\neq m'$ and repeat the same construction with $m'$ treated as the truth. Define
\[
L_{j,t}^{m,m'}:=\frac{p_{j,t}(Y_{j,t+1}\mid m,\cF_t)}{p_{j,t}(Y_{j,t+1}\mid m',\cF_t)}.
\]
For $\alpha\in[0,1]$, let
\[
\J_{j,t}^{m,m'}(\alpha):=-\log\!\int p_{j,t}(y\mid m',\cF_t)^{1-\alpha}p_{j,t}(y\mid m,\cF_t)^\alpha\,d\nu(y),
\]
and define
\begin{align}
\Info_t^{\rm gen}(m,m';s)&:=\sum_{\tau<t}\sum_j\J_{j,\tau}^{m,m'}(sv_j),\\
\underline{\Info}^{\rm gen}_t(s)&:=\min_{m\neq m'}\Info_t^{\rm gen}(m,m';s).
\label{eq:general-pairwise-information}
\end{align}
Conditional on $\cF_t$,
\[
\E_{m'}[(L_{j,t}^{m,m'})^\alpha\mid\cF_t]
=\exp[-\J_{j,t}^{m,m'}(\alpha)].
\]
If $|\log L_{j,t}^{m,m'}|\le b_j$ almost surely uniformly over all ordered pairs, histories, and selected actions, define the stationary geometric pairwise odds
\[
G_t^{m,m'}:=\prod_{i=1}^{N}
\left(\frac{\mu_{i,t}(m)}{\mu_{i,t}(m')}\right)^{v_i}.
\]
Then the corresponding pairwise process
\[
Z_t^{m,m'}(s)=\big(G_t^{m,m'}\big)^s
\exp\!\big(\Info_t^{\rm gen}(m,m';s)\big)
\]
is a nonnegative martingale and the same bounded network-disagreement argument applies. Replacing $\underline{\Info}_t(s)$ in the proof of Theorem~\ref{thm:anytime} by $\underline{\Info}_t^{\rm gen}(s)$ therefore gives the same Ville/union-bound guarantee. This establishes the stated general-observation anytime extension.

For the BSC query model, \eqref{eq:general-J} is zero when a query does not separate the pair and equals $\J_j(\alpha)$ when it does, recovering \eqref{eq:information}. This connects the certificate directly to the controlled-sensing formulation of action-dependent observation laws \citep{nitinawarat2013controlled}.

\section{Generalized Innovation Weights}
\label{app:reweight}
For completeness, consider
\[
    \widetilde\mu_{j,t+1}(m)
    \propto
    \mu_{j,t}(m)\ell_{j,t}(Y_{j,t+1}\mid m)^{\beta_j},
    \qquad \beta_j>0.
\]
The ratio recursion becomes
\[
    \rb_{t+1}=A\left(\rb_t+\boldsymbol\xi_t\right),
    \qquad \xi_{j,t}=\beta_j\log L_{j,t}.
\]
Every proof above is unchanged after replacing $\J_j(sv_j)$ by $\J_j(sv_j\beta_j)$, restricting $s$ to $sv_j\beta_j\le1$, and replacing $b_j$ in $\Delta_i$ by $\beta_jb_j$. With $\beta_j=1/v_j$, the leading certified information becomes $\sum_j\J_j(s)$ and is maximized at $s=1/2$. This inverse-stationary correction was introduced for fixed-observation decentralized detection by \citet{huang2025price}; Appendix~\ref{app:imbalance-proof} states its consequence for the adaptive-query certified exploration floor.

\section{Using the Certificate with Learned Query Policies}
The theorem can be used as a policy diagnostic without modifying a learned policy. In simulated or supervised evaluation, where the ground truth is known, log every chosen query and compute
\[
    \widehat\gamma_T(s)
    =\frac{1}{T}\min_{m\neq\mS}\Info_T(m;s).
\]
For a fixed, pre-specified $s$, a larger $\widehat\gamma_T(s)$ means that the policy has accumulated evidence more uniformly against its hardest competitor. If $s$ is optimized after the rollout, the resulting quantity is the post-hoc diagnostic score in \eqref{eq:empirical-certificate}, not an anytime-valid certificate by itself. Corollary~\ref{cor:stochastic-growth} shows what additional statement is needed to convert this diagnostic into an unconditional exponent guarantee: an exponentially decaying lower tail for the information process. This distinction is especially relevant for learned decentralized active-sensing policies such as those studied empirically in \citet{szostak2024deep}.

\section{Additional Experimental Details}
\label{app:experimental-details}

All experiments used CPU execution on a single 4-core workstation.

\paragraph{Rate fitting and plotted uncertainty.}
For 1D runs, let $e_t$ be network-average MAP error. We first retain $2\le t\le\max\{10,\lfloor0.85T\rfloor\}$ with $0.75/(n_{\rm trial}N)<e_t<0.8$; if fewer than five points remain, we fall back to positive points with $1\le t$ and $e_t<0.8$. We fit $\log e_t$ on the last contiguous retained segment and return n/a if fewer than three points remain. For 2D runs we retain $t\ge4$, $e_t>\max\{3/(n_{\rm trial}N),10^{-8}\}$, and $e_t\le0.30$; if fewer than five points remain, we use positive points with $t\ge4$ and $e_t<0.8$, again requiring five points.

All squared-loss curves use the squared distance of each agent's MAP estimate from the true target, averaged over agents within a trajectory and then over independent trajectories; they are not posterior Bayes risk. For 1D curves, pointwise intervals use Student-$t$ critical values (normal fallback); the 2D plotting code uses mean $\pm1.96$ standard errors. These intervals are computed on the linear scale. Log plots use small display floors only for zero means or nonpositive interval endpoints (e.g., $10^{-12}$ in the 1D MSE panel and $10^{-14}$ for the 2D band), so values below the Monte Carlo resolution are not interpreted as measured errors.

\paragraph{Graph-geometry study.}
Figure~\ref{fig:graph-convergence} uses $M=32$, $N=10$, homogeneous BSC crossover probability $\epsilon=0.25$, $5{,}000$ Monte Carlo trials, $80$ rounds, and exploration+bisection with one deterministic exploration round every five rounds. Fits use only positive-error points before the Monte Carlo floor.
\begin{center}
\begin{tabular}{lcc}
\toprule
Graph & Fitted exponent & Information score \\
\midrule
Complete & 0.4288 & 0.5380 \\
Ring     & 0.4035 & 0.5295 \\
Directed & 0.3384 & 0.5230 \\
\bottomrule
\end{tabular}
\end{center}

For completeness, Appendix~\ref{app:additional-policy-plots} repeats the graph-geometry experiment with posterior bisection, Thompson-style posterior sampling, and random thresholds. Their fitted exponent / information-score pairs are: posterior bisection---complete $0.5283/0.6758$, ring n/a$/0.6648$, directed $0.4194/0.6566$; Thompson---complete $0.3412/0.5780$, ring $0.3176/0.5726$, directed $0.3146/0.5707$; random---complete $0.0558/0.0417$, ring $0.0541/0.0417$, directed $0.0516/0.0415$. ``n/a'' denotes a run for which the predefined positive-error pre-floor fitting rule did not return a stable exponent.

\begin{figure*}[t]
\centering
\begin{subfigure}{0.49\textwidth}
  \includegraphics[width=\linewidth]{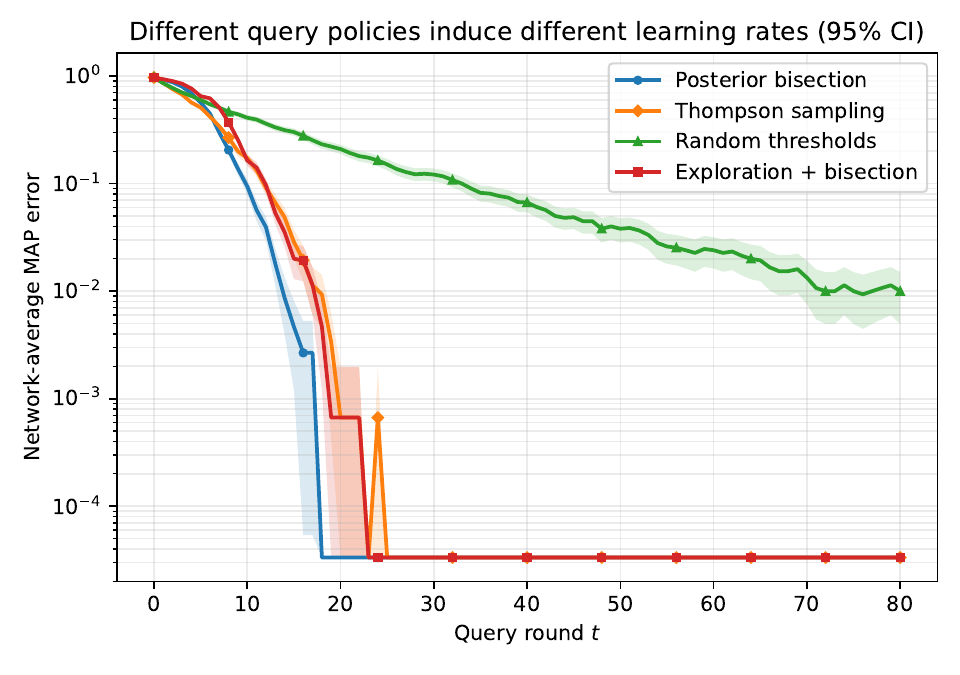}
  \caption{Query-policy comparison on the common 1D problem.}
  \label{fig:policy-certificate}
\end{subfigure}\hfill
\begin{subfigure}{0.49\textwidth}
  \includegraphics[width=\linewidth]{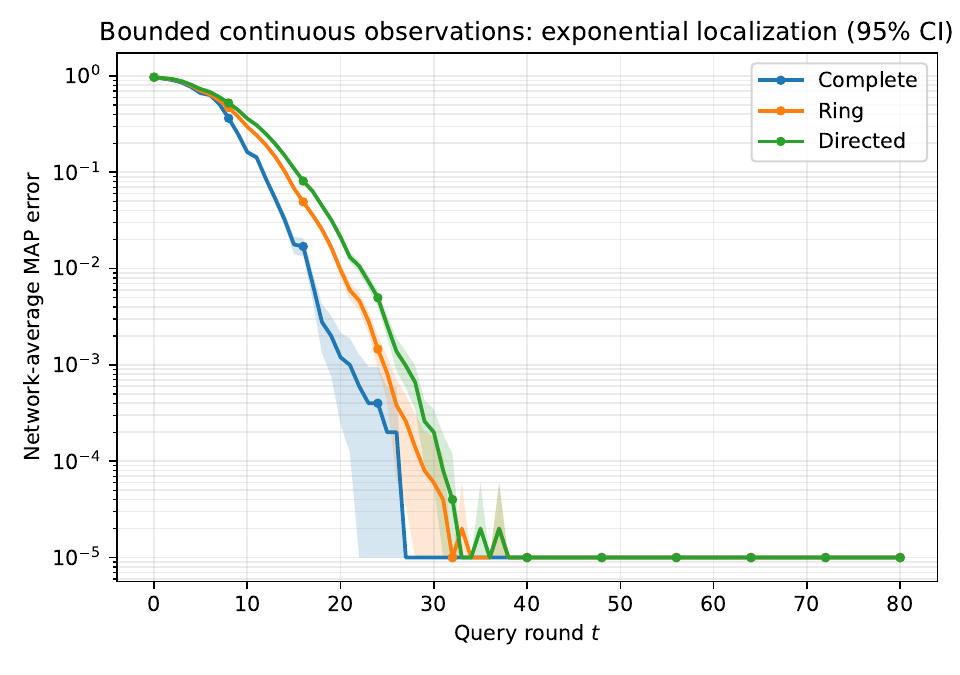}
  \caption{Information-matched bounded continuous sensor.}
  \label{fig:continuous-sensor}
\end{subfigure}
\caption{\textbf{Additional theory-facing checks.} Left: four query policies on the common 1D problem. Right: the bounded continuous-observation model, which exercises the general controlled-observation construction.}
\label{fig:appendix-sanity}
\end{figure*}

\paragraph{Policy comparison, Fig.~\ref{fig:policy-certificate}.}
The common setting is $M=32$, $N=10$, homogeneous $\epsilon=0.25$, $1{,}500$ trials, $80$ rounds, and the complete graph. For reproducibility, the Thompson-style rule samples $\widetilde m_{j,t}\sim\mu_{j,t}$ and chooses
\[
k_{j,t}^{\mathrm{TS}}\in\argmax_{1\le k\le M-1}\sum_{m=1}^{M}\mu_{j,t}(m)\,\1\{q_k(m)\neq q_k(\widetilde m_{j,t})\},
\]
with random tie-breaking.
\begin{center}
\begin{tabular}{lcc}
\toprule
Policy & Fitted exponent & Information score \\
\midrule
Posterior bisection   & 0.4487 & 0.6756 \\
Thompson-style sampling & n/a & 0.5773 \\
Random thresholds     & 0.0580 & 0.0413 \\
Exploration+bisection & 0.4418 & 0.5378 \\
\bottomrule
\end{tabular}
\end{center}
The Thompson curve reaches the empirical floor before the fitting rule identifies a stable pre-floor linear region, so we do not assign it a panel-(a) exponent.

\paragraph{Pooled bottleneck study, Fig.~\ref{fig:proxy-comparison}.}
The ten-seed study spans four policies, three communication graphs, and homogeneous/heterogeneous sensor profiles, giving $240$ configurations; $210$ satisfy the deterministic rate-fit rule above. The other $30$ receive n/a because no valid pre-floor fitting segment is available; no exponent is imputed for them. Correlations and regressions are therefore explicitly conditional on the $210$-configuration subset rather than the full design. All correlations and regression quantities use $10{,}000$ seed-cluster-bootstrap replicates with whole seed blocks resampled jointly. The numerical results are reported once in Section~\ref{sec:exp1d} and annotated in Fig.~\ref{fig:proxy-comparison}.

\begin{figure}[t]
\centering
\includegraphics[width=0.49\columnwidth]{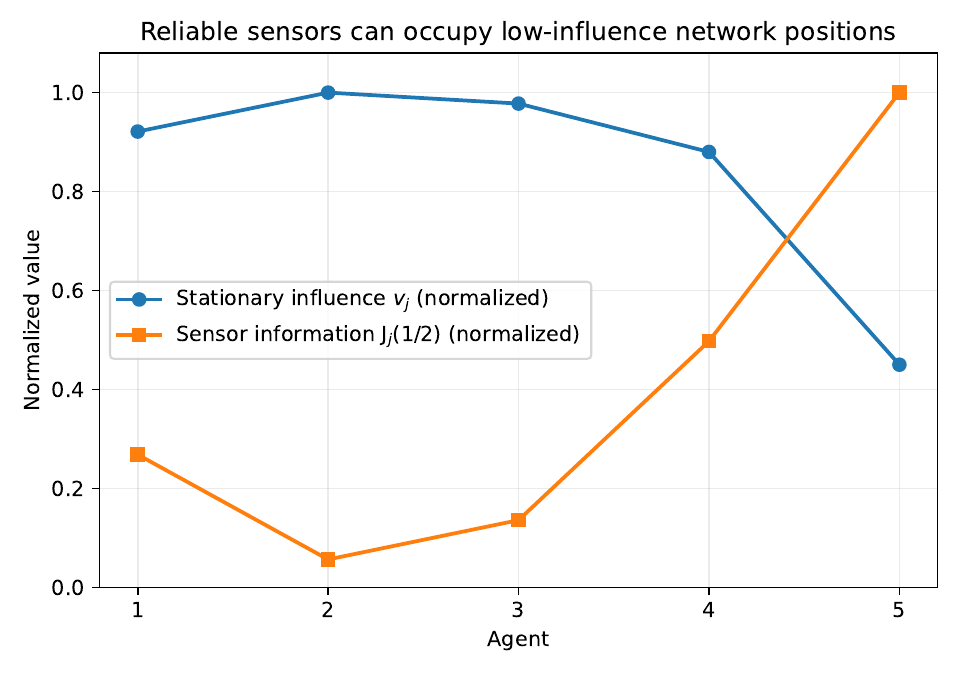}
\caption{\textbf{Directed stationary-influence stress test.} Sensor information is deliberately anti-aligned with stationary network influence, exposing the coupling represented by $\J_j(sv_j)$.}
\label{fig:directed-stress}
\end{figure}

\paragraph{Directed heterogeneous-sensor study.}
Figure~\ref{fig:directed-stress} uses the deliberately imbalanced five-agent directed graph, $10{,}000$ Monte Carlo trials, $100$ rounds, and exploration+bisection. The stationary distribution is
\[
(0.2178,\,0.2364,\,0.2312,\,0.2081,\,0.1065),
\]
the BSC crossover probabilities are
\[
(0.20,\,0.35,\,0.275,\,0.125,\,0.05),
\]
and the corresponding midpoint Chernoff/Bhattacharyya information values \citep{bhattacharyya1943measure} are
\[
(0.2231,\,0.0472,\,0.1131,\,0.4133,\,0.8304).
\]
For exploration+bisection, the fitted observable-tail exponents are $0.4637$ (standard update) and $0.4633$ (inverse-stationary update), with raw post-hoc information-score values $0.5407$ and $0.1590$. We do not compare these post-hoc score magnitudes directly because innovation reweighting changes the fractional-moment parameterization.

The additional policy variants in Appendix~\ref{app:additional-policy-plots} give the following exponent / information-score pairs for standard versus inverse-stationary updates: posterior bisection $(\text{n/a}/0.6040)$ versus $(\text{n/a}/0.1594)$; Thompson $0.3770/0.4429$ versus $0.5061/0.1283$; random $0.0478/0.0386$ versus $0.0502/0.0435$; and exploration+bisection $0.4637/0.5407$ versus $0.4633/0.1590$. These policy-dependent finite-horizon outcomes are why the main text treats Experiment~3 as a mechanism stress test rather than evidence that inverse-stationary weighting uniformly improves empirical convergence speed.

\paragraph{Worst-competitor versus average-pair proxy.}
For every valid pooled configuration, the simulation records both the truth-indexed worst-competitor diagnostic score $\widehat\gamma_T^{\rm diag}$ from \eqref{eq:empirical-certificate} and the all-pairs aggregate rate in \eqref{eq:average-pair-proxy}. Both correlations are computed on the identical $210$ configurations. Resampling whole seed blocks preserves the paired design and avoids treating the multiple policy--graph--sensor configurations associated with one seed as independent bootstrap units; Section~\ref{sec:exp1d} reports the resulting statistics.

\paragraph{Continuous-observation study.}
Conditional on query bit $b\in\{0,1\}$, the external-validity experiment uses
\begin{equation}
 p_j(y\mid b)=\frac{\kappa_j}{2\sinh\kappa_j}\exp\!\big((2b-1)\kappa_j y\big),
 \qquad -1\le y\le1,
 \label{eq:continuous-sensor}
\end{equation}
whose log-likelihood ratio obeys $|\log[p_j(y\mid1)/p_j(y\mid0)]|\le2\kappa_j$. For a separating query,
\begin{equation}
\J_j^{\rm cont}(\alpha)
=-\log\!\left[\frac{\sinh((2\alpha-1)\kappa_j)}{(2\alpha-1)\sinh\kappa_j}\right],
\label{eq:continuous-J}
\end{equation}
with continuous limit $\J_j^{\rm cont}(1/2)=\log(\sinh\kappa_j/\kappa_j)$. We choose $\kappa_j$ numerically so that
\begin{equation}
\J_j^{\rm cont}(1/2)=-\log\!\bigl(2\sqrt{\epsilon_j(1-\epsilon_j)}\bigr),
\label{eq:continuous-match}
\end{equation}
matching the midpoint information of the reference BSC. The study uses $M=32$, $N=10$, $5{,}000$ trials, $80$ rounds, exploration+bisection with one deterministic exploration round every five rounds, and the same complete/ring/directed graph constructions as Fig.~\ref{fig:graph-convergence}. The simulator uses the exact likelihood in the Bayes update and the exact increment in \eqref{eq:continuous-J}. The fitted exponent / worst-score pairs are complete $0.4248/0.5372$, ring n/a$/0.5285$, and directed n/a$/0.5221$; the latter two reach zero Monte Carlo error before the predefined positive-error fitting rule identifies a stable pre-floor linear region.

\paragraph{Structured 2D spatial-query experiment.}
We use an $8\times8$ grid ($M=64$), $N=10$, $100$ rounds, and a $42$-query library of vertical, horizontal, and $\pm45^\circ$ half-planes. The certified family comprises the $14$ axial cuts. On the complete graph with homogeneous $\epsilon=0.25$ and $1{,}500$ trials per policy, the exponent / worst-score / average-pair-score triples are posterior bisection $0.7571/0.5510/0.5245$, Thompson-style sampling $0.2953/0.4767/0.4857$, random spatial queries $0.0914/0.0928/0.4393$, and exploration+bisection $0.4660/0.4673/0.5275$. All final empirical MAP errors are zero at round $100$.

The pooled 2D bottleneck study is run with $500$ trials per configuration over four policies, three graphs, homogeneous/heterogeneous sensor profiles, and ten seed blocks ($240$ configurations); the heterogeneous profile spans BSC crossover probabilities $0.13$--$0.35$. All $240$ configurations yield valid pre-floor fits. With $10{,}000$ paired seed-cluster-bootstrap replicates, $r_{\rm worst}=0.768$ (95\% CI $[0.761,0.777]$), $r_{\rm avg}=0.484$ (95\% CI $[0.474,0.494]$), $\Delta r=0.285$ (95\% CI $[0.270,0.299]$), and the worst-score regression slope is $0.773$ (95\% CI $[0.736,0.811]$).

\begin{figure*}[t]
\centering
\begin{subfigure}{0.49\textwidth}
  \includegraphics[width=\linewidth]{figures_2d/fig2d_policy_map_error_ci.pdf}
  \caption{Network-average MAP error.}
\end{subfigure}
\begin{subfigure}{0.49\textwidth}
  \includegraphics[width=\linewidth]{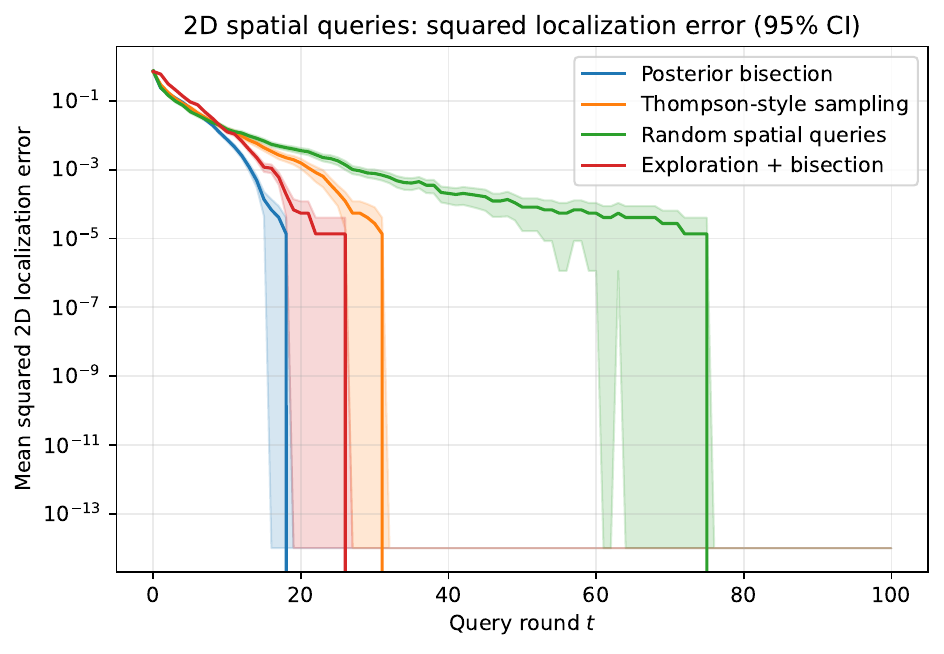}
  \caption{Squared Euclidean localization loss.}
\end{subfigure}
\caption{\textbf{Structured 2D localization curves.} Four query policies on the $8\times8$ spatial grid; shading denotes pointwise 95\% normal-approximation confidence intervals over independent trajectories.}
\label{fig:appendix-2d-curves}
\end{figure*}

\paragraph{Tail fitting.}
The exact 1D/2D fitting rules are stated at the start of Appendix~\ref{app:experimental-details}; zero-error rounds are never fitted. Resolving deeper tails would require more independent trials or a dedicated rare-event estimator, so display-floor values are not treated as observations.

\section{Additional Query-Policy Experiments}
\label{app:additional-policy-plots}
This appendix collects companion localization-loss plots, additional query-policy curves, the bounded continuous-observation study, and directed-influence robustness checks. The results are mechanism checks rather than claims of policy optimality.

\begin{figure*}[t]
\centering
\begin{subfigure}{0.49\textwidth}
  \includegraphics[width=\linewidth]{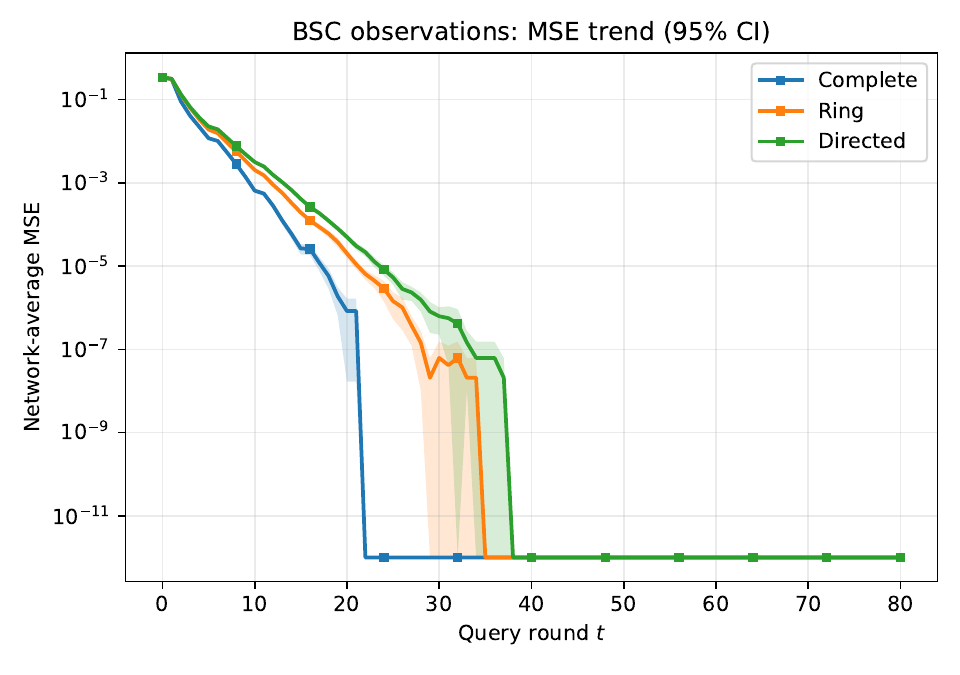}
  \caption{Exploration+bisection: MSE (companion to Fig.~\ref{fig:graph-convergence}).}
\end{subfigure}
\begin{subfigure}{0.49\textwidth}
  \includegraphics[width=\linewidth]{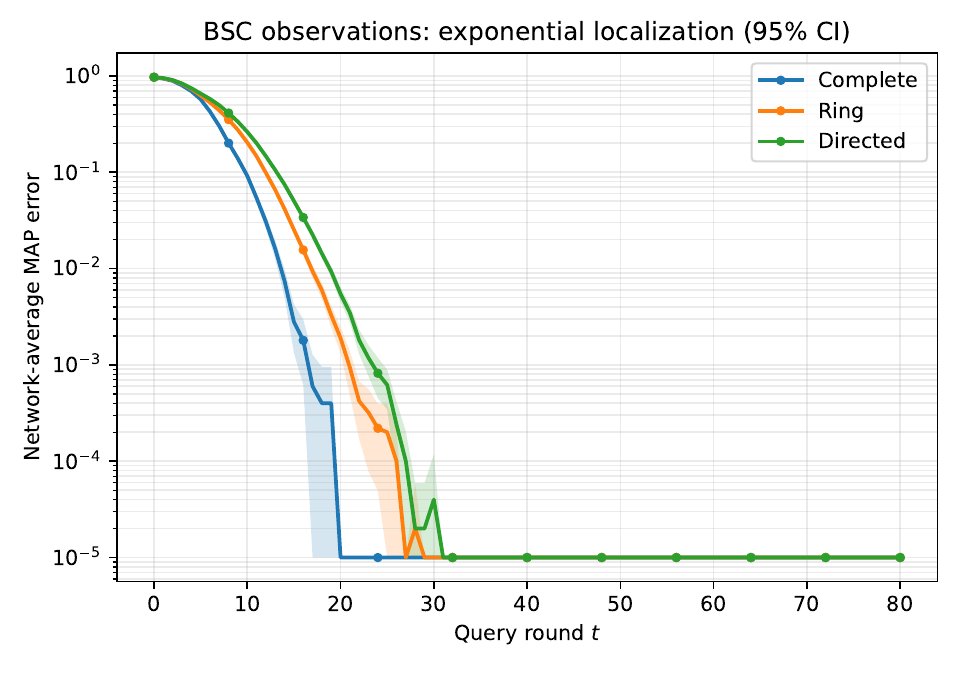}
  \caption{Posterior bisection: MAP error.}
\end{subfigure}

\begin{subfigure}{0.49\textwidth}
  \includegraphics[width=\linewidth]{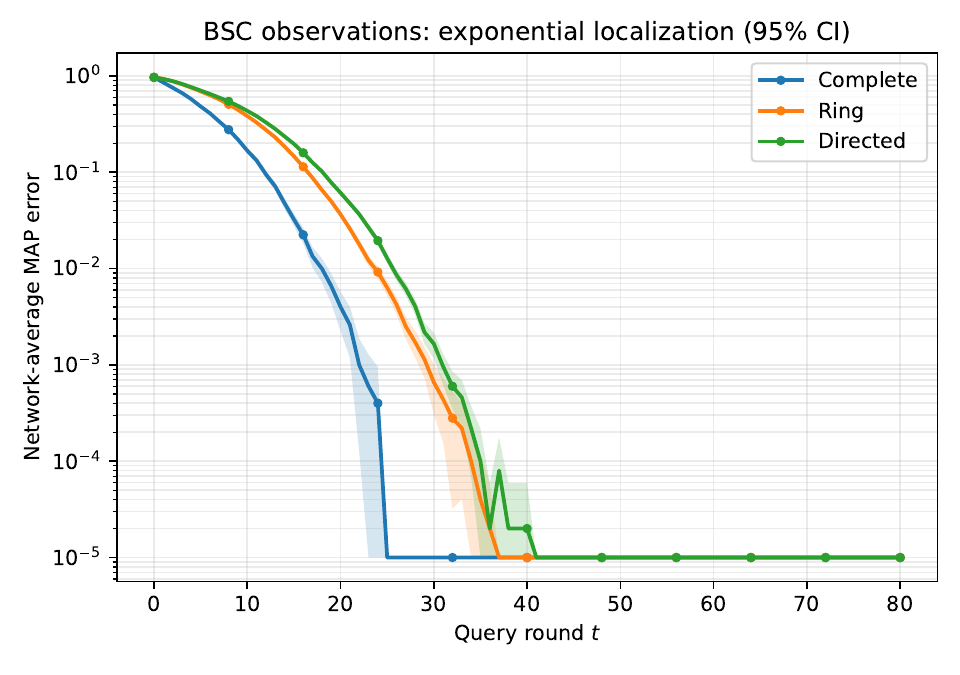}
  \caption{Thompson-style sampling: MAP error.}
\end{subfigure}
\begin{subfigure}{0.49\textwidth}
  \includegraphics[width=\linewidth]{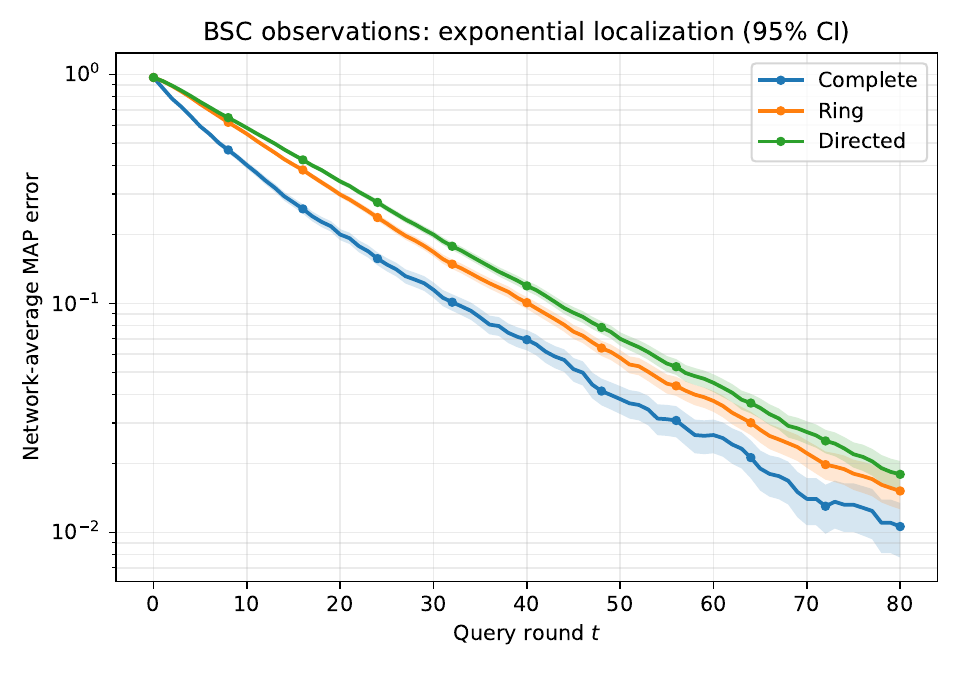}
  \caption{Random thresholds: MAP error.}
\end{subfigure}
\caption{\textbf{Graph-geometry robustness and MSE companion.} The MSE panel for the main exploration+bisection study and MAP-error results for the remaining query policies. Full MAP/MSE outputs are included with the released figure scripts.}
\label{fig:appendix-graph-policies}
\end{figure*}

\begin{figure*}[t]
\centering
\begin{subfigure}{0.49\textwidth}
  \includegraphics[width=\linewidth]{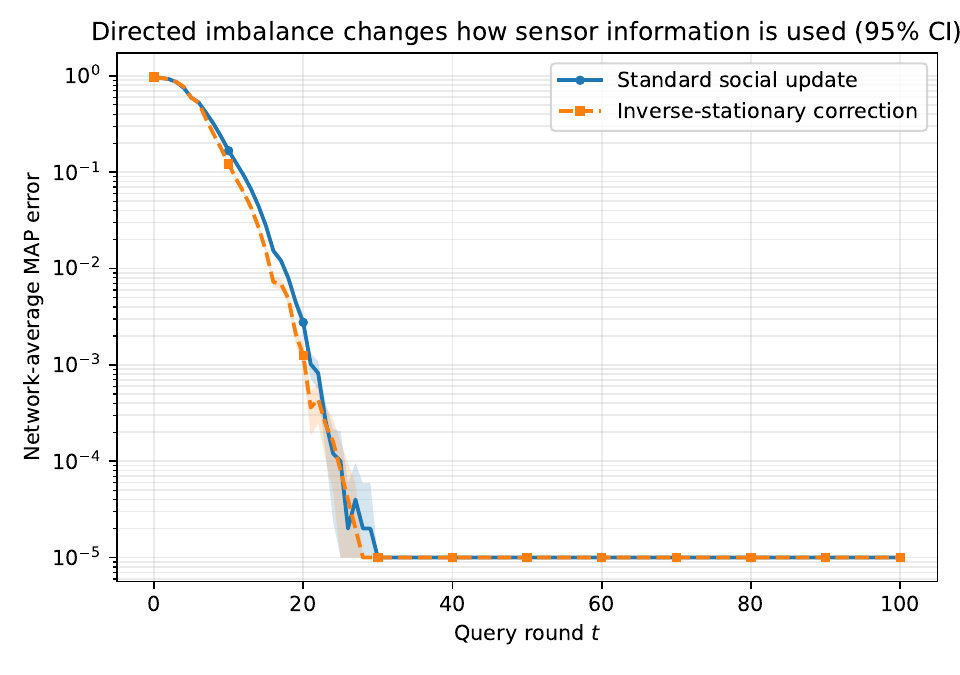}
  \caption{Exploration+bisection.}
\end{subfigure}
\begin{subfigure}{0.49\textwidth}
  \includegraphics[width=\linewidth]{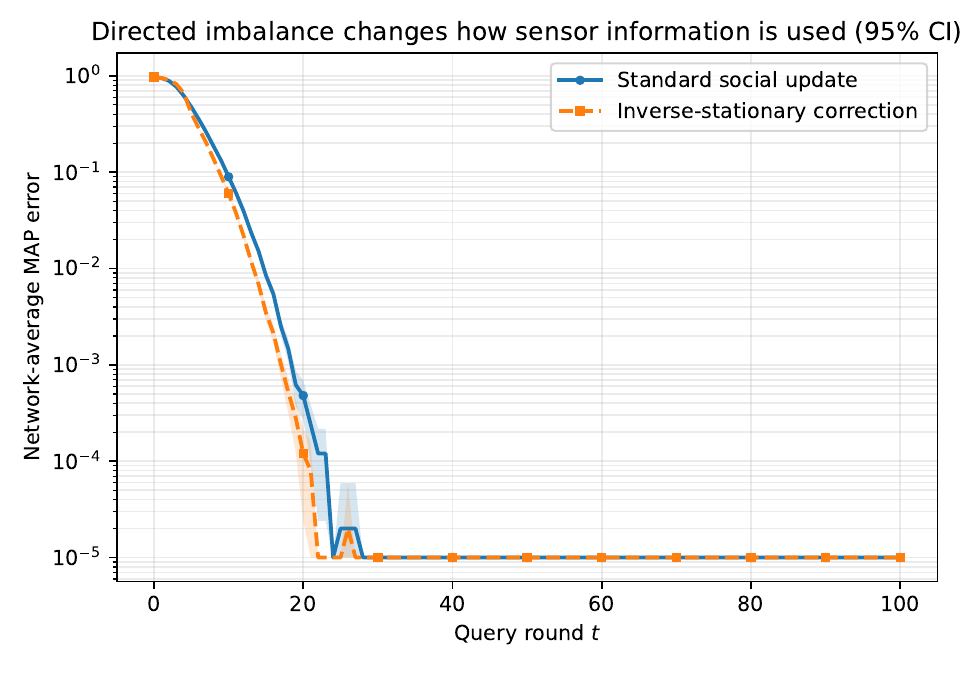}
  \caption{Posterior bisection.}
\end{subfigure}

\begin{subfigure}{0.49\textwidth}
  \includegraphics[width=\linewidth]{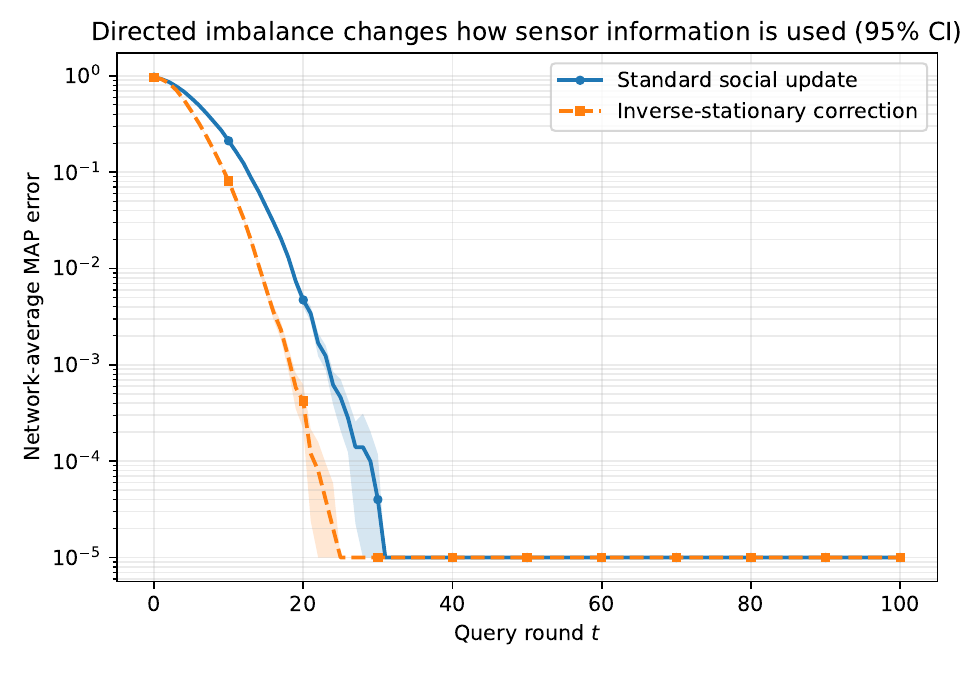}
  \caption{Thompson-style sampling.}
\end{subfigure}
\begin{subfigure}{0.49\textwidth}
  \includegraphics[width=\linewidth]{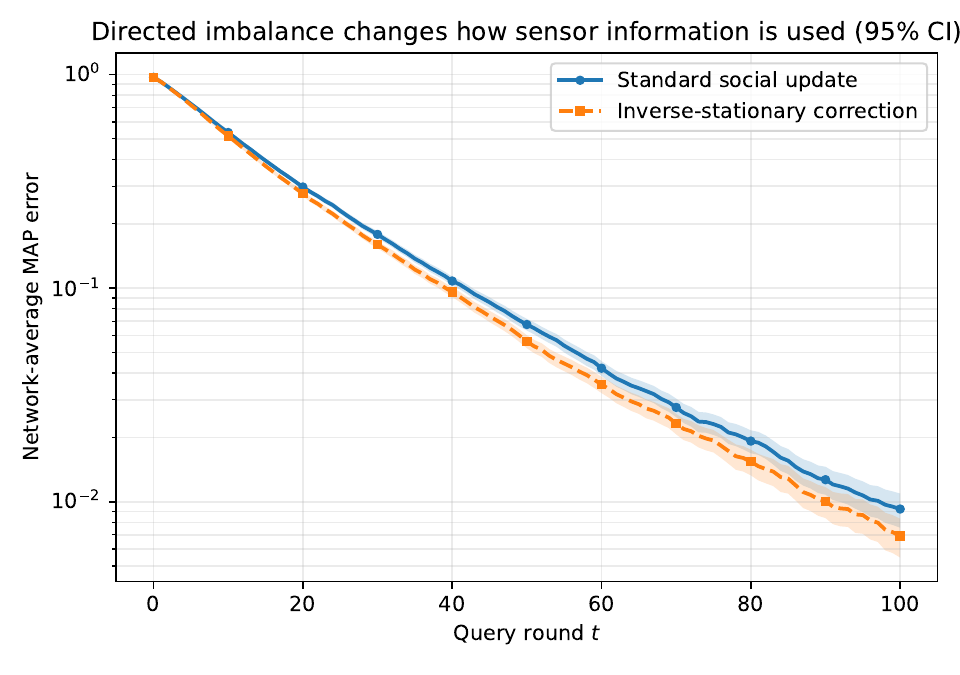}
  \caption{Random thresholds.}
\end{subfigure}
\caption{\textbf{Directed-network update comparisons.} Standard versus inverse-stationary updating under four query policies, all using the same anti-aligned sensor-quality/stationary-influence profile shown in Fig.~\ref{fig:directed-stress}.}
\label{fig:appendix-directed-policies}
\end{figure*}

\end{document}